\documentclass[manuscript]{aastex}
\usepackage{color}
\usepackage{subfigure}

\def\gtrsim{\lower 2pt \hbox{$\, \buildrel {\scriptstyle >}\over
{\scriptstyle \sim}\,$}}
\def\lesssim{\lower 2pt \hbox{$\, \buildrel {\scriptstyle <}\over
{\scriptstyle \sim}\,$}}

\def\suzaku{{\sl Suzaku}}

\def\suzaku{{\sl Suzaku}}

\shorttitle{}
\shortauthors{}

\begin{document}

\title{Fe Line Diagnostics of Cataclysmic Variables and Galactic Ridge X-ray Emission}

\author{Xiao-jie Xu}
\affil{School of Astronomy and Space Science and Key Laboratory of Modern Astronomy and Astrophysics, Nanjing University, Nanjing, P. R. China 210093}
\author{Q. Daniel Wang}
\affil{Department of Astronomy, University of Massachusetts Amherst, MA 21003, U.S.A.}
\author{Xiang-Dong Li}
\affil{School of Astronomy and Space Science and Key Laboratory of Modern Astronomy and Astrophysics, Nanjing University, Nanjing, P. R. China 210093}

\begin{abstract}
The properties of the Galactic Ridge X-ray Emission (GRXE) observed in the 2-10 keV band
place fundamental constraints on various types of X-ray sources in the Milky Way.
Although the primarily discrete origin of the emission is now well established, the responsible populations of these sources remain uncertain, especially at relatively low fluxes.
To provide insights into this issue, we systematically characterize the 
Fe emission line properties of the candidate types of the sources in the solar neighborhood
and compare them with those measured for the GRXE. Our source sample includes 6 symbiotic stars (SSs), 16 intermediate polars (IPs), 3 polars, 16 quiescent dwarf novae (DNe) and 4 active binaries (ABs); they are all observed with the \textit{Suzaku} X-ray Observatory. The data of about 1/4 of these sources are analyzed for the first time.
We find that the mean equivalent width ($EW_{6.7}$) of the 6.7-keV line and the mean 7.0/6.7-keV line ratio are $107\pm16.0$ eV and $0.71\pm 0.04$ for intermediate polars and 
$221\pm 135$ eV and $0.44\pm 0.14$ for polars, respectively, which are all substantially different from
those ($490\pm15 $~eV and $0.2\pm 0.08$) for the GRXE. Instead, the GRXE values 
are better agreed by the $EW_{6.7}$ ($438\pm 84.6$~eV) and the ratio ($0.27\pm 0.06$) observed for the DNe. 
 We further find that the $EW_{6.7}$ is strongly correlated with the 2-10-keV luminosity of 
the DNe, which can be characterized by the relation $EW_{6.7}=(438\pm95 {\rm~eV}) (L/10^{31}~{\rm~ergs~s^{-1} })^{(-0.31\pm0.15)}$. Accounting for this correlation, 
the agreement can be improved further, especially when the contributions from 
other classes sources to the GRXE are considered, which all have low $EW_{6.7}$ values.
We conclude that the GRXE mostly consists of typically faint, but numerous DNe, 
plus ABs, while magnetic CVs are probably mainly the high-flux representatives 
of the responsible populations and dominate the GRXE only in harder energy bands.

\end{abstract}


\section{Introduction}

Although discovered more than 30 years ago, the exact origin of the Galactic Ridge X-ray Emission (GRXE) observed in the 2-10 keV range remains largely uncertain \citep[e.g.,][]{Wor82,rev09,hon12}. 
The overall spectral shape 
of the GRXE, with a total 2-10 keV luminosity of $\sim (1-2) \times 10^{38}{\rm~erg~s^{-1}}$, can be approximately characterized as a two-temperature optically thin thermal plasma with $kT_{1}\sim 1$ keV and 
$kT_{2}\sim 15$  keV 
over the  2 - 50~keV (or $kT_{2}\sim 6$ keV over 2-10 keV ) range covered by the
\textit{Suzaku}~\citep[e.g.,][]{yua10,yua12}. 
The spectrum of the GRXE also contains
 emission lines; the three most prominent ones are from Fe K$\alpha$
transitions: 6.4-keV line from neutral or weakly-ionized species, 6.7-keV and 7.0-keV lines from Helium-like and Hydrogen-like ions (e.g., \citealt{uch13}). 
Both the line and continuum properties  of the GRXE provide important clues about
its origin.

Two scenarios have been proposed to explain the GRXE. The first scenario assumes that the GRXE arises primarily from truly diffuse hot plasma in and near the Galactic disk, while the second proposes that the emission represents a superposition of numerous point-like sources, including their light scattered by the interstellar medium (ISM)~\citep{molaro14}. The 1~Ms \textit{Chandra} deep exposure carried out by \citet{rev09} toward the `Limiting Window' in the Galactic bulge successfully resolved out about 80\% of the GRXE (at $\sim 6$~keV)  into individual sources with a detection limit down to the 2-10 keV luminosity of $\sim$ $10^{30}{\rm~erg~s^{-1}}$ at the Galactic center distance of 8~kpc (see also \citealt{hon12}). Furthermore, the spatial distribution of the GRXE
closely follows that of the stellar mass in the Galaxy~\citep{rev06}. These
results strongly support the point source scenario, leaving little room 
for a substantial contribution from the diffuse hot plasma, probably except for  the Galactic center region ($\lesssim 0.5^\circ$ from Sgr A*; e.g., \citealt{uch11}). 

However, the nature of the point sources  responsible for the GRXE remains elusive. It has been suggested that the dominating source population might be cataclysmic variables (CVs), or more specifically, magnetic ones 
(mCVs), plus a minor contribution from active binaries \citep[ABs; e.g.,][]{rev09}. CVs are binaries, each consisting a white dwarf (WD) accreting matter from a main-sequence or red-giant companion. Based on the magnetic field strengths of the WDs, CVs are classified into magnetic [including polars and intermediate polars (IPs)] or non-magnetic ones [including novae, quiescent dwarf novae (DNe), symbiotic star (SSs), etc.]. Comparatively less luminous are ABs, binaries of normal 
stars whose X-ray emission is enhanced by strong magnetic field 
due to spin-orbital coupling. The luminosity functions (LFs) of nearby CVs and ABs were studied by \citet{saz06,byc10,pre12}. Based on their LFs, as well as 
spectral and temporal behaviors~\citep{rev09,hon12}, relative bright point 
sources (with the 2-10~keV luminosity $L_{\rm X} \gtrsim$  $ 10^{32}{\rm~erg~s^{-1}}$) observed in the GRXE are mostly considered to be mCVs. 
The nature of fainter sources, especially those below $ 10^{31}{\rm~erg~s^{-1}}$ is much less clear. Individually they 
are typically too faint to allow spectral and temporal 
analyses with existing X-ray data. Collectively, the LF of 
such sources suggests that they contribute more to the GRXE than those 
bright ones; the sources in the range of 
$L_{\rm X} \sim$ $10^{30-31}{\rm~erg~s^{-1}}$, for example, are about 30 times more 
in numbers than the bright ones and account for
75\% of the resolved emission in the 6-8~keV range~\citep{rev09}. The 
LF may, however, vary significantly from the the stellar disk, to the bulge, and to the center of the Galaxy~\citep[e.g.,][]{Perez15}. So the consideration of the local LF alone cannot be conclusive with regard to the nature of the sources in the GRXE. Therefore, an improved understanding of the GRXE will be interesting not only for its 
own right, but for providing fundamental constraints on the integrated properties of 
various X-ray source populations in the Galaxy as well. These constraints 
are also essential to the correct estimation and subtraction of the source 
contributions and hence to the study of diffuse hot plasma in other galaxies~\citep[e.g.,][]{lizy07}.

\begin{figure}
\centering
\includegraphics[width=0.45\textwidth,keepaspectratio=true,clip=true]{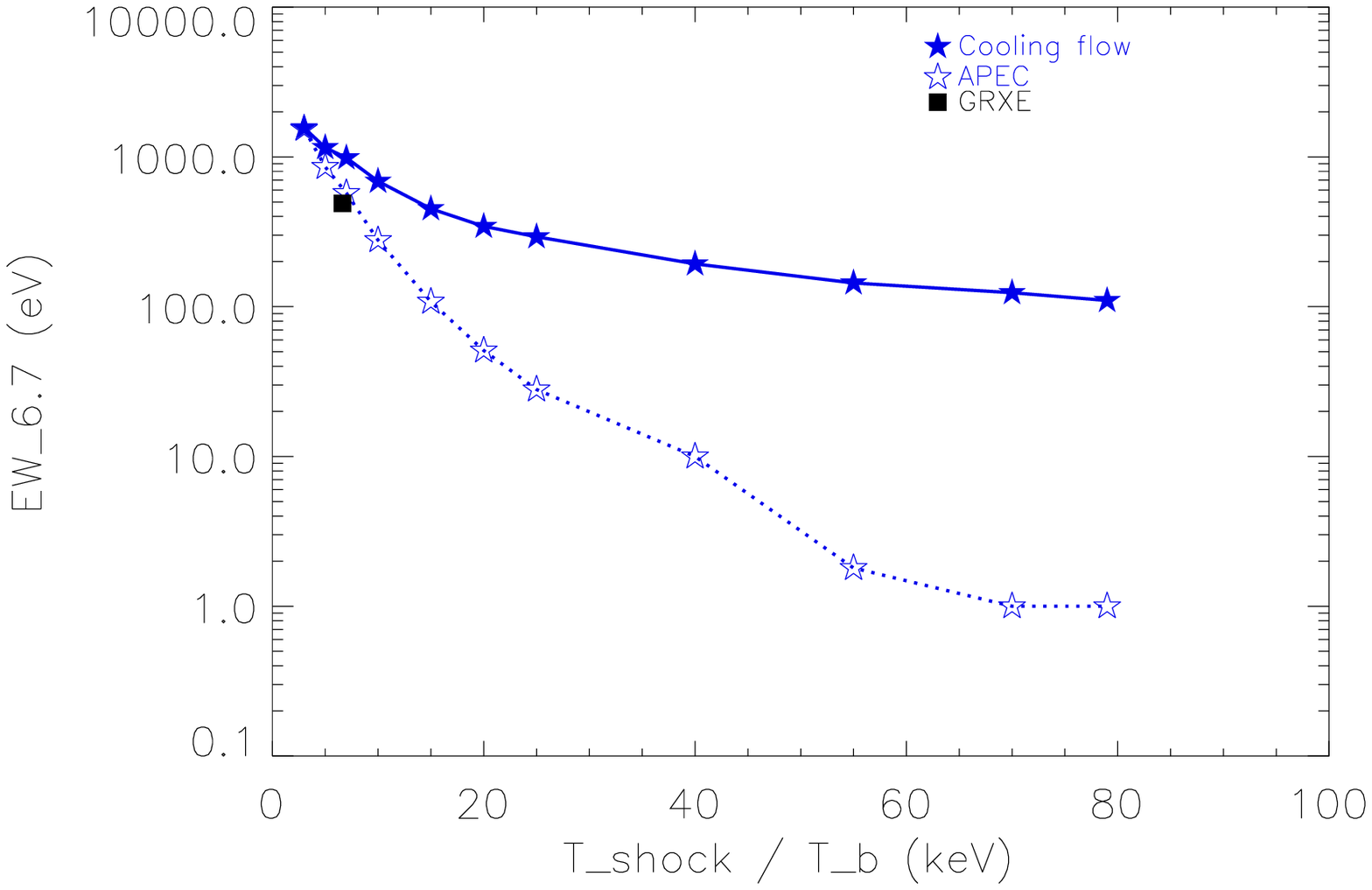}
 \includegraphics[width=0.45\textwidth,keepaspectratio=true,clip=true]{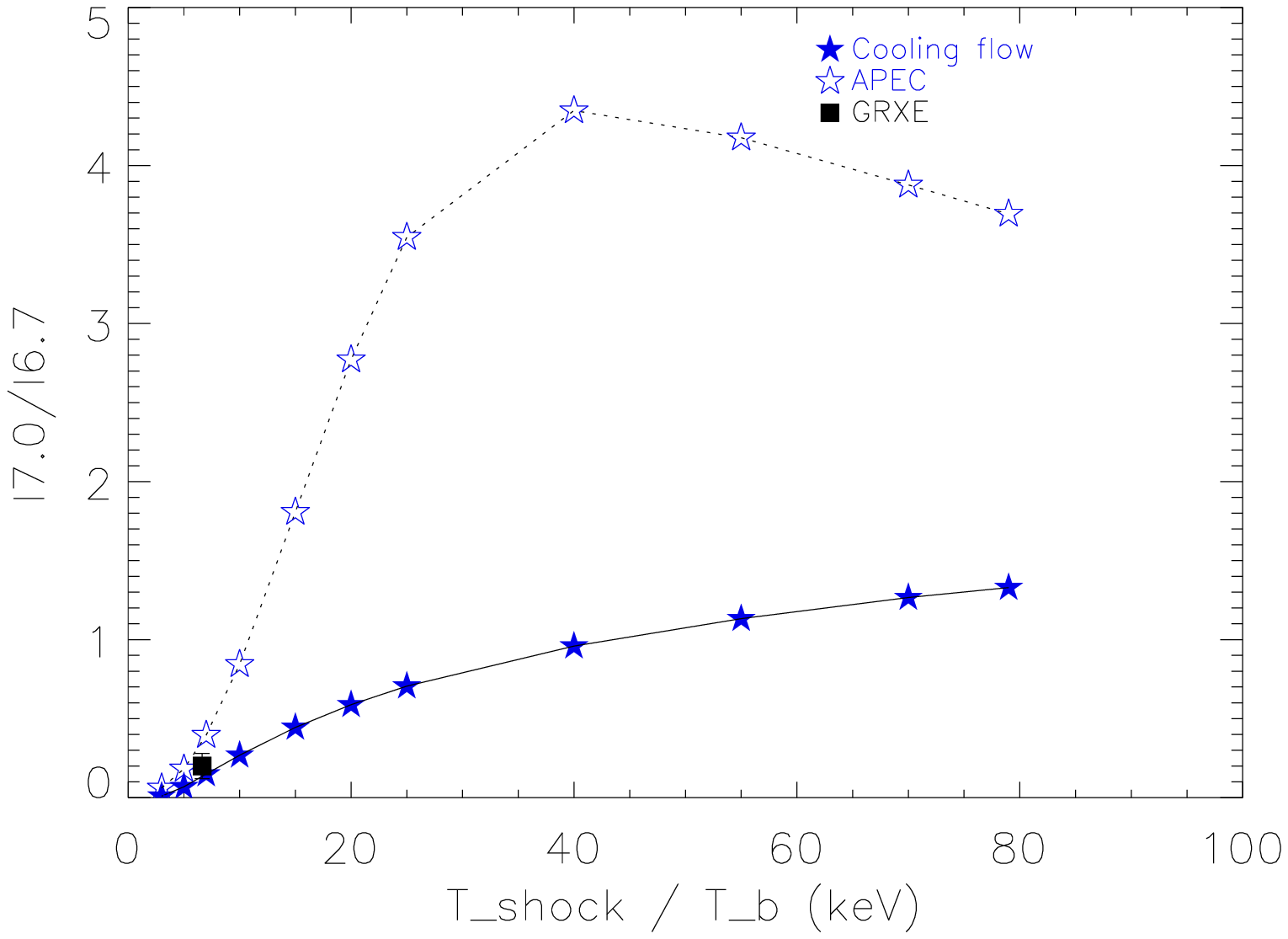}
\caption{$EW_{6.7}$ (left panel) and $I_{7.0}/I_{6.7}$ (right panel) as function of the plasma or shock temperature for two spectral models. $EW_{6.7}$ is proportional to the Fe abundances (the solar value is assumed here; \citealt[e.g.,][]{gray08}).  The measured $T_{\rm b}$, $EW_{6.7}$ and $I_{7.0}/I_{6.7}$ ~\citep{uch13} for the GRXE are marked for reference. 
}
   \label{Fig:01}
\end{figure}

We here focus on the use of the diagnostic Fe K emission lines to 
probe the nature of the faint sources in the GRXE in the 2-10~keV band. To do so, we first 
characterize the mean intensities of the emission lines from  local sources
of various known classes (mCVs, DNe, ABs, SSs, and polars). 
Specifically, both the EW of the 6.7-keV ($EW_{6.7}$) and the flux ratio of 
the 7.0 to 6.7~keV lines ($I_{7.0}/I_{6.7}$) are sensitive probers 
of hot plasma temperature ($T$; Fig.~\ref{Fig:01}). $EW_{6.7}$ always decreases, while 
$I_{7.0}/I_{6.7}$ increases with increasing $T$ in the 1-20~keV range. 
In an mCV, an accretion disk is not expected 
(at least near the WD); the gravitational potential energy of the infall gas (e.g., in an accretion column)
is converted into the heat almost entirely at a standing shock near 
the surface of the WD~\citep[e.g.,][]{lon11}. In contrast, for a non-magnetic CV, 
a cool accretion disk is expected, which radiates about half of the gravitational energy and 
hardly contributes to the X-ray emission.
The other half of the energy, retained in the Keplerian motion at the inner 
boundary of the disk, can effectively heat the gas when eventually falling onto 
the WD through the boundary layer. The maximum temperature that the gas can reach depends on the exact heating mechanism (e.g., via a single shock or a series of weaker ones), but should 
be about a factor of  $\gtrsim 2$ smaller than that for 
an mCV of same WD mass. 
This expectation is consistent with existing observations, which  
show that mCVs generally have smaller $EW_{6.7}$ 
and greater $I_{7.0}/I_{6.7}$ than non-magnetic CVs~\citep[e.g.,][]{ish99,ezu99,anz09,muk09,ish09}.
Moreover, unlike the EWs of the Fe lines, $I_{7.0}/I_{6.7}$ is independent of 
either the metallicity of the hot plasma or the continuum emission, which can be strongly affected by X-ray absorption and/or reflection. Thus $I_{7.0}/I_{6.7}$ provides the most reliable diagnostic of the plasma temperature in the 1-20~keV range. Interestingly, the typical temperature of GRXE falls in this range\citep{yua12,uch13}, making Fe lines sensitive probes in comparisons among different classes of CVs (with typical temperatures below 20~keV) or with the GRXE. 

Additional constraint on the plasma temperature can be obtained from the measurement of the 6.4-keV emission line,
which is predominantly from the fluorescence of ionizing 
photons ($> 7.12$~keV) by cold irons.
The strength of the line, relative to the other lines or to the continuum,
is sensitive to both the temperature of the hot plasma, which produces the photons, and
the amount and geometry of the surrounding cool materials. In a CV, for example,
about half of the photons are typically intercepted by 
the WD surface, while additional fluorescence may
occur in the accretion column (for an mCV),  the accretion disk 
(non-magnetic CV), or the stellar envelope (SS). For the GRXE, however, the 6.4-keV 
line emission can be partly diffuse in origin, e.g., due to the 
fluorescence in dense ISM clouds or to their interaction with low-energy cosmic 
rays. Therefore, the GRXE can be used to place an upper limit to the overall 
population of 6.4-keV line-emitting sources in the Galaxy.

Various measurements of the Fe lines exist for local CVs, as well as for the GRXE.
Probably the most extensive measurements for the GRXE are made by \citet{uch13}, who show
that the EWs of the lines change systematically from the center to the ridge of the Galaxy. Toward the ridge, $EW_{6.4}= 110 \pm10$ eV, $EW_{6.7} = 490 \pm15$ eV, $EW_{7.0} = 110 \pm 10$ eV and $I_{7.0}/I_{6.7}=0.2\pm0.08$; the errors of these EWs values were not given in \citet{uch13} and are estimated from \citet{yua12}, based on similar counting statistics of the lines, although the latter work was based on data collected for the Galactic bulge (specifically, observations taken in the Galactic longitude and latitude ranges $|l| \lesssim 3$ and $|b| \lesssim 2$, excluding the very central region of $|l| \lesssim 1$ and $|b| \lesssim 0.5$).
The relative intensities of the lines change systematically from the ridge to the center; in particular, $I_{7.0}/I_{6.7} \approx 0.2$ in the ridge increases to $\sim  0.25$ in the budge, to $\sim 0.4$ in the center of the Galaxy.  
The $EW_{6.4}$ value varies strongly with position, especially in the central region, and 
sometimes with time~\citep[e.g.,][]{yua12,muno07}.  The $EW_{6.7}$ ($I_{7.0}/I_{6.7}$) values of the ridge appear to be significantly higher (lower) than those of relatively bright (but unidentified) sources (above $\sim 5 \times 10^{32} {\rm~ergs~s^{{-1}}~cm^{-2}}$) detected by XMM \citep{war14} and those of local mCVs \citep[e.g.,][]{ish99,uch13}. Understanding this apparent inconsistency is a motivation of the work reported here. 

We first systematically analyze the Fe emission line properties of local CVs and ABs and then confront the results with the measurements of the GRXE.
The existing analyses of local CVs and ABs are rather inhomogeneous, based on data collected with various X-ray telescopes, such as \textit{ASCA, Suzaku, Chandra HETG} and on different spectral modeling procedures. The sample sizes of the individual analyses are also small (typically less than 20 sources)~\citep[e.g.,][]{ezu99,hel04,bas05,ran06,sch14}.  Our sample includes 41 CVs and 4 ABs, all observed with  \textit{Suzaku}. The results from our analysis of this sample can then directly be compared with similar \suzaku\ measurements of the GRXE (with minimum biases), providing new insights into the nature of the responsible source populations. 
The rest of the paper is organized as follows: In \S~2 we describe our sample selection and data analysis methods; We present our results in \S~3; We compare them with existing results and discuss the implications in \S~4; And finally, in \S~5, we provide a summary of the work.

\section{Sample Selection \& Data Analysis}

\textit{Suzaku} operated between 2005 and 2015. It had four X-ray Imaging Spectrometers (XIS): Three of them were made of front illuminated CCDs (XIS-0, -2 and -3), while
the other of back-illuminated one (XIS-1). These spectrometers had the best spectral 
resolution ($E/\delta E \sim 20$ to $50$) among the similar instruments that have operated in the 0.3-10 keV range. This high resolution is essential to the analysis reported here.
The moderate spatial resolution of the instruments is also suited for
local CVs and ABs: even though they are relatively bright sources, 
pile-ups are not an issue.

We cross-correlate the \textit{Suzaku} online archive with \citet{rit03}'s CV catalog and \citet{eke08}'s AB catalog to search for available observations. We find 48 publicly available observations on 41  CVs (including 6 SSs, 16 IPs, 3 polars and 16 DNe; multiple observations on several sources) and 4 observations on 4 ABs. The observation log of this source sample
is presented in Table~\ref{tbl-01}. One of the DNe, SS Cyg, was undergoing outburst during the observation. Nevertheless, it shows spectral properties similar to those of other DNe and is thus included in our subsequent analysis.

We reprocess the data downloaded from the \textit{Suzaku} archive, using the software package heasoft (version 12.8.1;
\citealt{arn96}). Briefly, the event files are reduced with the standard 
pipeline routine {\sl aepipeline} and the latest calibration files (XIS:20150312). For each observation,
we  extract an on-source spectrum and an off-source background 
spectrum, together with the response matrix (rmf) and effective area (arf) files,
using {\sl xselect}. 
The on-source spectrum is from a circular region 
with a typical radius $R=200\arcsec$; however, it is reduced to $R=120\arcsec$ to $150\arcsec$ for 20 observations, in which the sources are close to CCD edges.
The background spectrum is from an annulus with the inner 
and outer radii equal to $\sim 250\arcsec$ and $400\arcsec$, except for the edges. Our 
results are not sensitive to the exact selection of the background area (which is always greater 
than the on-source region), because the sources are all quite bright.

We conduct the spectral analysis, using the xspec software package. 
The spectra from all XIS chips are jointly fitted to improve the counting statistics.
Our focus is on the Fe emission lines and on the spectral shape of the continuum in
the 5-10 keV range, which minimizes potential complications due to the 
thermal emission from the WD surfaces, as well as the absorption by the cold and warm gases 
at lower energies. Also in this range, the confusion from the emission from diffuse
interstellar gas in the galactic disk is negligible~\citep[e.g.,][]{yua12}. 
We model the continuum with the optically thin thermal plasma model {\sl apec}
(with metallicity set to zero) for ease of comparison with previous results on  GRXE\citep[e.g.,][]{yam09,yua12,uch13}. This modeling also gives a characteristic plasma temperature ($T_{b}$) for each 
source. We emphasize that $T_{b}$ here is just used to characterize the hardness
of the spectral continuum. The shock temperature $T_{\rm shock}$ in the cooling flow model (as used in Fig.~\ref{Fig:01}) would be
a more physical quantity comparing to $T_{b}$ (e.g., when explaining $I_{\rm 7.0}/I_{\rm 6.7}$-$EW_{\rm 6.7}$ anti-correlation in Fig. 4, see next section), but $T_{\rm shock}$ were not used to model the spectra of GRXE in previous works\citep[e.g.,][]{yam09,yua12,uch13} and thus is not used in this work. A detailed analysis of the spectra of sampled sources with the cooling flow model is in preparation. 
We here mostly use $EW_{\rm 6.7}$ and $I_{\rm 7.0}/I_{\rm 6.7}$ as two independent line diagnostics of the plasma temperature. Accordingly,
we construct a spectral model consisting of three Gaussians, representing the 6.4-keV, 6.7-keV and 7.0-keV lines. The parameters of this model include the centroid energies, widths, and (relative) intensities ($I_{6.4}/I_{6.7}$, $I_{6.7}$, and $I_{7.0}/I_{6.7}$) of the lines. Building the 
desirable relative intensities of the 
lines into the model (instead of using the standard normalizations of the individual 
Gaussians) automatically accounts for their correlations in error measurements. Additionally, the unabsorbed 2-10 keV flux and luminosity $L_{\rm X}$ of each source are calculated; the latter includes its error in the distance measurements (see Table 1).

For each class of our sample sources, we calculate the means and intrinsic dispersions 
of each individual parameter ($T_{\rm b}$, $EW_{6.4}$, $EW_{6.7}$, $EW_{\rm 7.0}$, and
$I_{7.0}/I_{6.7}$). For each class, we may construct an expected $\chi^2$ statistic as
\begin{equation}
  \chi^2=\sum_{i=1}^{n} \frac{(x_{\rm i}-a)^2}{b^2+s_{\rm i}^2},
\end{equation}
where $n$ is the total source number, $x_i$ and $s_i$ are the measured
value and error of the i-th source, whereas $a$ and $b$ are the expected mean and standard deviation of the parameter (assuming a normal distribution). The $\chi^2$ minimization relative to
$a$ gives the equation:
\begin{equation}
  a = \sum_{i=1}^{n} \frac{x_{\rm i}}{(b^2+s_{\rm i}^2)} / \sum_{i=1}^{n} \frac{1}{(b^2+s_{\rm i}^2)}.
\end{equation}
This equation, together with setting $\chi^2$ to its expected value $n-2$ (the number
of degrees of freedom), allows us to solve for $a$ and $b$. The error of $a$ can be further calculated from
\begin{equation}
  \sigma^2=\frac{1}{\sum_{i=1}^{n} [1/(s_{\rm i}^2+b^2)]}.
\end{equation}

\section{Results}
\begin{figure}
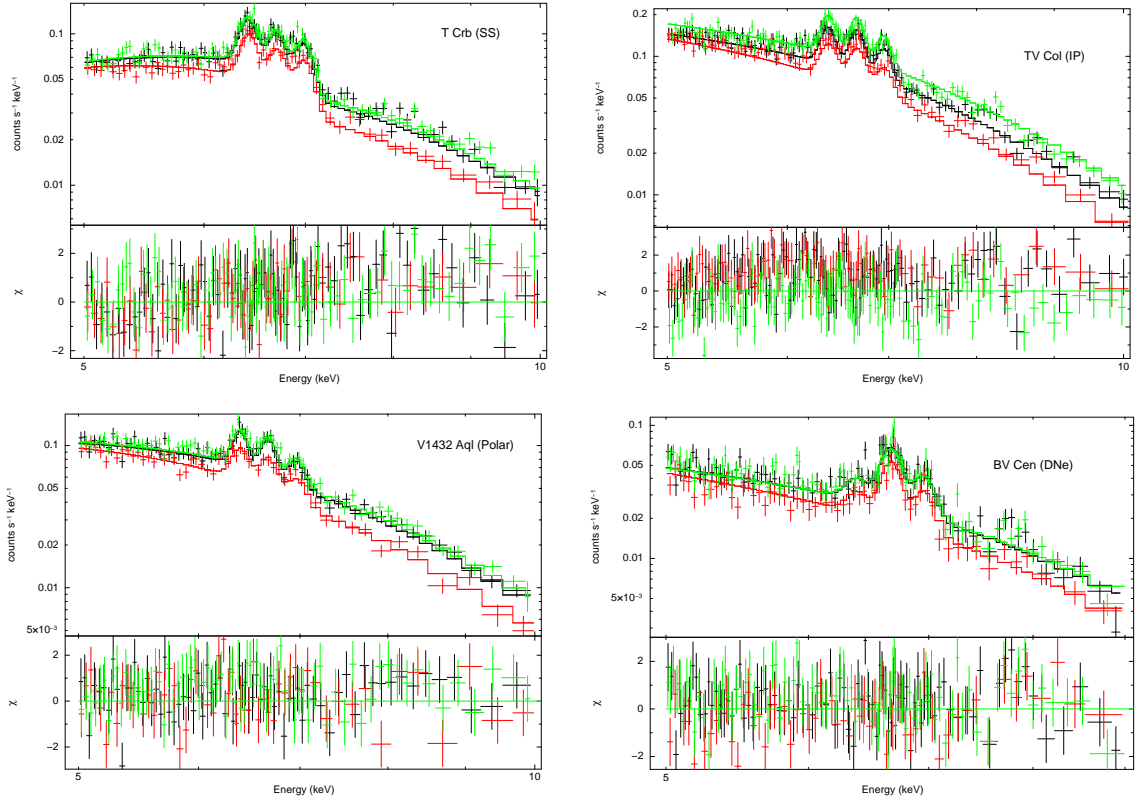

\centering
\subfigure{\includegraphics[height=3in,width=2in,angle=270]{tcrb.ps}}
\subfigure{\includegraphics[height=3in,width=2in,angle=270]{tvcol.ps}}
\\
\subfigure{\includegraphics[height=3in,width=2in,angle=270]{v1432.ps}}
\subfigure{\includegraphics[height=3in,width=2in,angle=270]{bvcen.ps}}
\renewcommand{\figurename}{good}
   \caption{Examples of CV spectra with the best-fit model of an absorbed apec (with the metallicity set to zero) plus three Gaussians for the Fe lines. The black-, red- and green-colored data points represent spectra from XIS-0, -1, and -3 chips. Upper left - T Crb (SS); upper right - TV Col (IP); lower left - V1432 Aql (polar); lower right - BV Cen (DN). Spectra are rebinned for plotting only.}
   \label{Fig:02}
\end{figure}

\begin{figure}
\centering
\includegraphics[height=4in,width=6in]{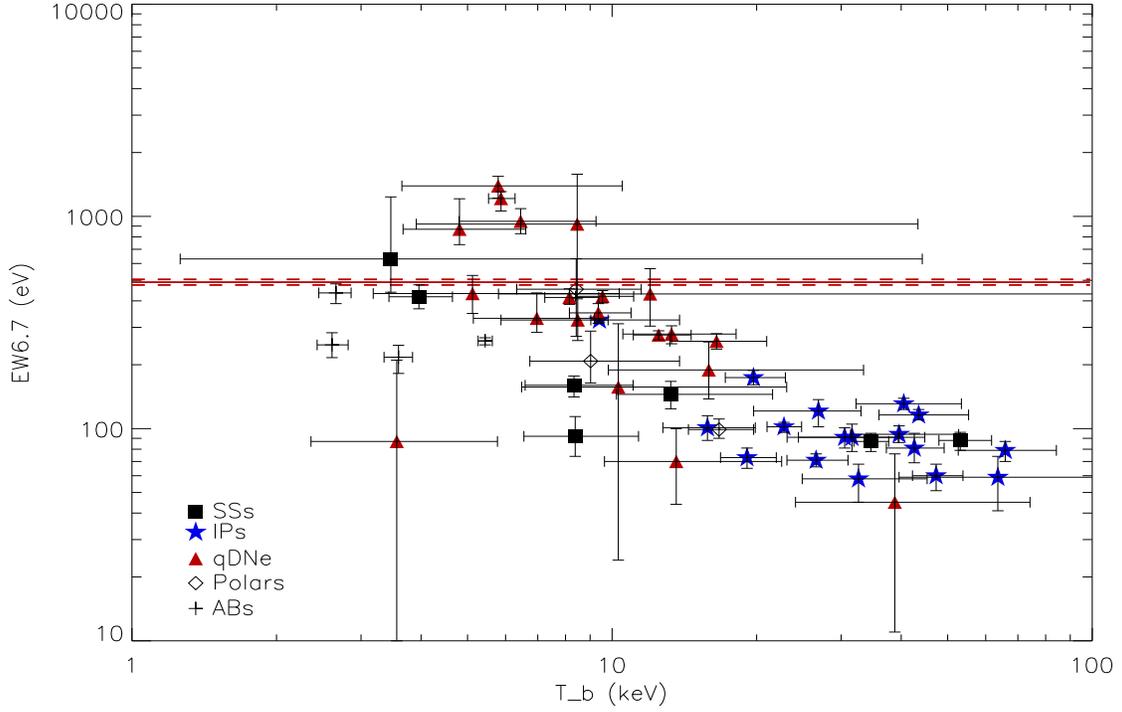}
   \caption{$EW_{6.7}$ vs. $T_{\rm b}$ of our sample sources. Individual classes are labeled differently. The solid and dashed red lines mark the $EW_{6.7}$ value and its $1\sigma$ error range of the GRXE. Fig~\ref{Fig:03} to Fig~\ref{Fig:07} includes multiple measurements for sources with multiple observations (see Table 2 for details): SS73-17 (two measurements), XY Ari (two measurements), VW Hyi (four measurements). }
   \label{Fig:03}
\end{figure}

Our main results are summarized in Table 2 and Fig~\ref{Fig:02} to Fig~\ref{Fig:08}. Fig~\ref{Fig:03} to Fig~\ref{Fig:07} includes multiple measurements for sources with multiple observations (see Table~\ref{tbl-02} for details): SS73-17 (two measurements), XY Ari (two measurements) and VW Hyi (four measurements). Fig~\ref{Fig:02} shows the spectra, together with the best-fitted models, 
for four sources as an example for each CV class considered here. 
The most distinct differences among the spectra are the Fe line 
ratios: large $I_{7.0}/I_{6.7}$  and $I_{6.4}/I_{6.7}$ for mCVs (polars and 
IPs), large $EW_{6.7}$ for the DNe, and exceptionally enhanced 6.4-keV line 
intensity for the SSs. In general, the model fitting is acceptable, 
judged from the $\chi^2$ values. Tables~\ref{tbl-02} and \ref{tbl-03} summarize our results, while following figures illustrate trends and correlations among the measured parameters. 

Fig~\ref{Fig:03} shows an apparent anti-correlation between $EW_{6.7}$ and $T_{\rm b}$, which holds even for individual classes (though with lower significance). Classwise, IPs  have the highest $T_{\rm b}$ (all above $\sim 15$~keV), while  DNe have relatively low $T_{\rm b}$ (mostly below 15~keV); the mean $T_{\rm b}$ ratio is about  3, consistent with the expectation from the different accretion (and hence heating) processes between the two classes (\S~1). The polars tend to have lower temperatures, compared to IPs, which is also expected (due to the contribution from relatively soft cyclotron line emission; \citealt{lon11}).  Interestingly, none of the IPs  has $EW_{6.7}$ as high as the value measured for the GRXE! While a considerable fraction of DNe in our sample also have  similarly small $EW_{6.7}$ (but with large error bars), the rest seem to be consistent with, or even significantly larger than, the GRXE value; these latter  DNe 
tend to have low $T_{b}$. Not surprisingly,  ABs tend to have the lowest $T_{\rm b}$. But they are not necessarily high in $EW_{6.7}$ due to their relatively low metallicity \citep[e.g.,][]{ber98,mal98,fra01}

\begin{figure}
\centering
\includegraphics[height=4in,width=6in]{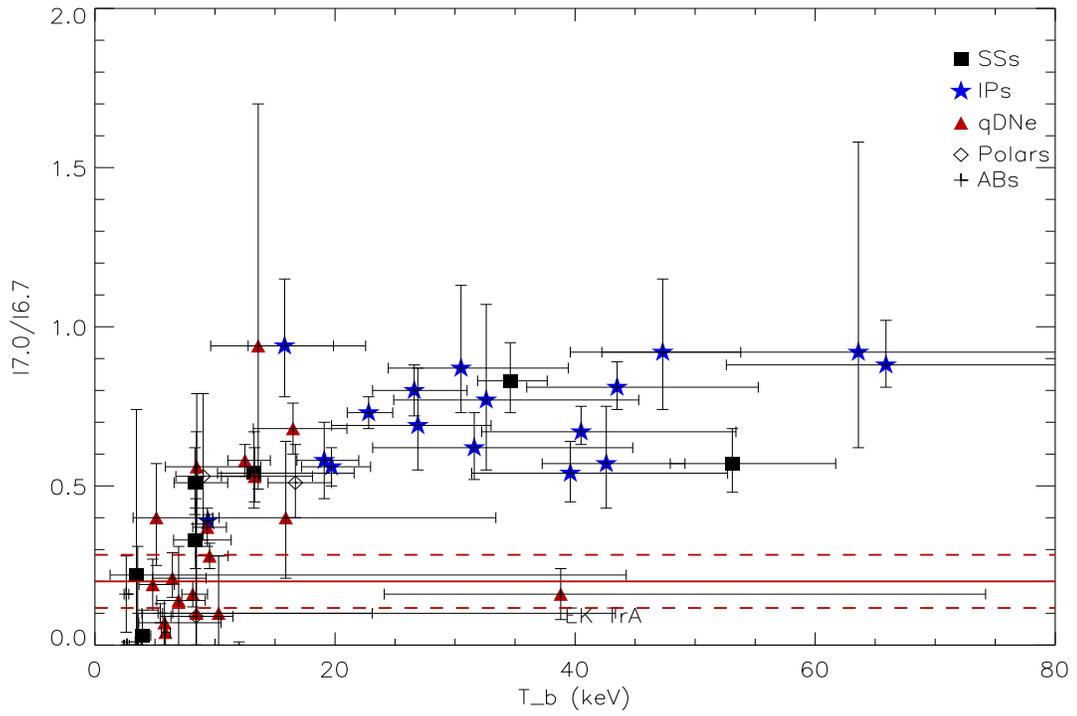}
   \caption{$I_{7.0}/I_{6.7}$ vs. $T_{\rm b}$ of our sample sources. The solid and dashed red lines mark the $I_{7.0}/I_{6.7}$ value and  $1\sigma$ error range of the GRXE. The labeled abnormal DN, EK TrA, is discussed in the text. }
   \label{Fig:04}
\end{figure}

\begin{figure}
\centering
\includegraphics[height=4in,width=6in]{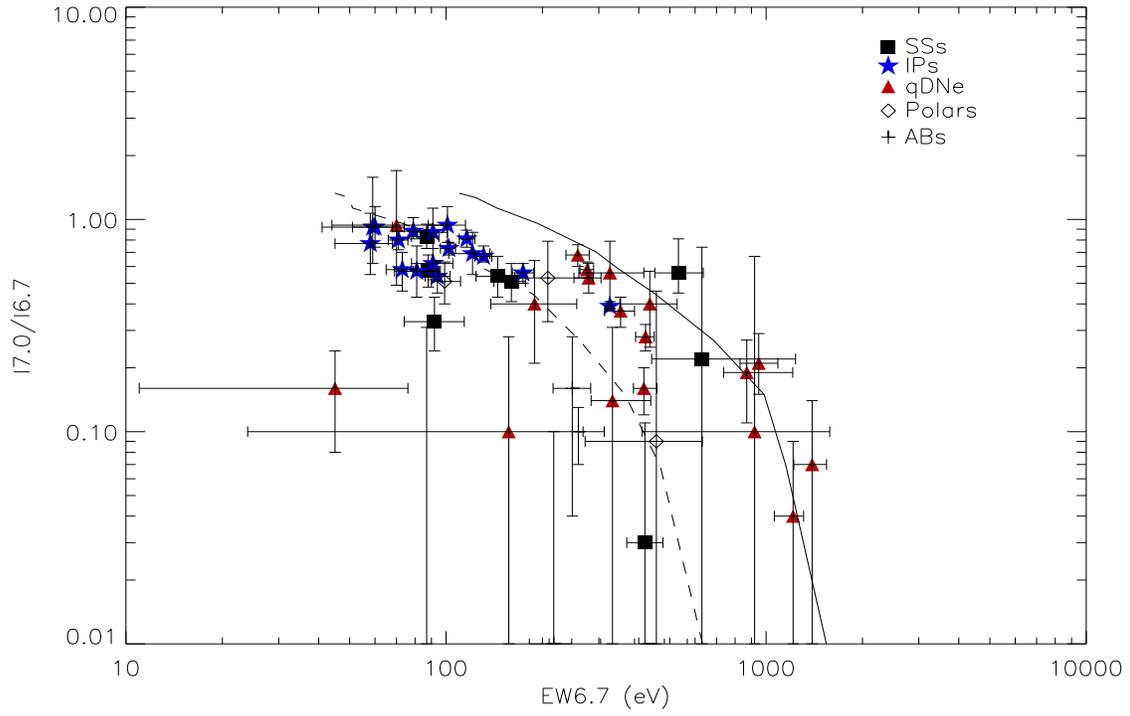}
   \caption{$I_{7.0}/I_{6.7}$ vs. $EW_{6.7}$ of our sample sources, compared with the values predicted by cooling flow model with  metallicity $Z=Z_{\odot}$ (solid line) and $Z=0.3Z_{\odot}$ *dashed line), respectively. }
   \label{Fig:05}
\end{figure}

Fig.~\ref{Fig:04} shows $I_{7.0}/I_{6.7}$ vs. $T_{\rm b}$ of our sample sources. The correlation between these two parameters, as well as the $I_{7.0}/I_{6.7}$ values of individual classes, is consistent with the trends seen in Fig~\ref{Fig:03}. Probably the only significant exception is 
EK TrA (a DN), which shows a low $I_{7.0}/I_{6.7}$, but a seemly high $T_b$, deviating from the overall correlation. But the 
ratio is consistent with the values of other DNe. As stated in \S~1, the ratio is typically 
a more reliable temperature diagnostic than $T_{\rm b}$ or $EW_{6.7}$. If this is the case, then we may speculate the continuum of the source may be somewhat enhanced and hardened (to explain its high $T_{b}$ and possibly slightly low $EW_{6.7}$ values, unusual for a DN; Fig~\ref{Fig:03}). Again, none of the IPs are consistent with the GRXE in terms of the $I_{7.0}/I_{6.7}$ ratio.  In general, IPs have higher $T_{\rm b}$ and higher $I_{7.0}/I_{6.7}$ values than  DNe and ABs. The higher $T_{\rm b}$ values of IPs can be naturally explained by greater expected in-falling velocities at the stand shocks and therefore higher shock temperatures  than in non-magnetic CVs (e.g., \S~1; see also~\citealt{lon11}). A higher temperature also naturally leads to a larger Hydrogen-like Fe ion fraction, and therefore a higher $I_{7.0}/I_{6.7}$ value. The spectra of our sampled ABs generally do not show significant 7.0-keV lines, thus their $I_{7.0}/I_{6.7}$ are less than $0.2$ (see Fig~\ref{Fig:04} and Fig~\ref{Fig:05} for details). As shown in Fig~\ref{Fig:05}, the dependence of $I_{7.0}/I_{6.7}$ on $EW_{6.7}$ is consistent with the cooling flow model predictions.

\begin{figure}
\centering
\includegraphics[height=4in,width=6in]{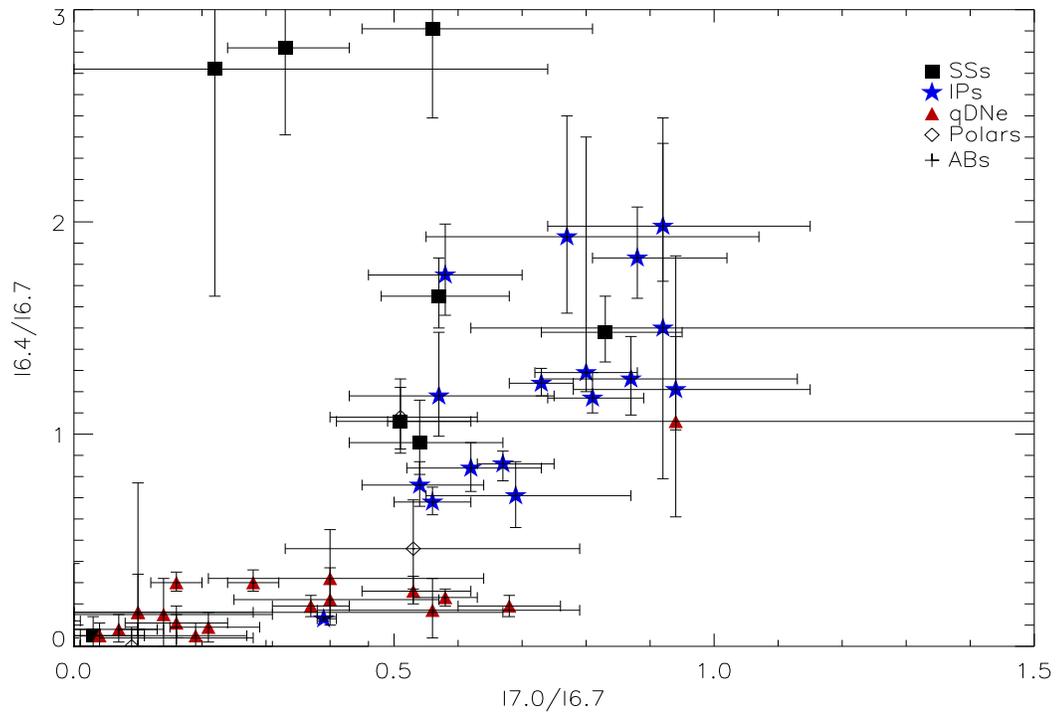}
   \caption{$I_{6.4}/I_{6.7}$ vs. $I_{7.0}/I_{6.7}$ of our sample sources.}
   \label{Fig:06}
\end{figure}

\begin{figure}
\centering
\includegraphics[height=4in,width=6in]{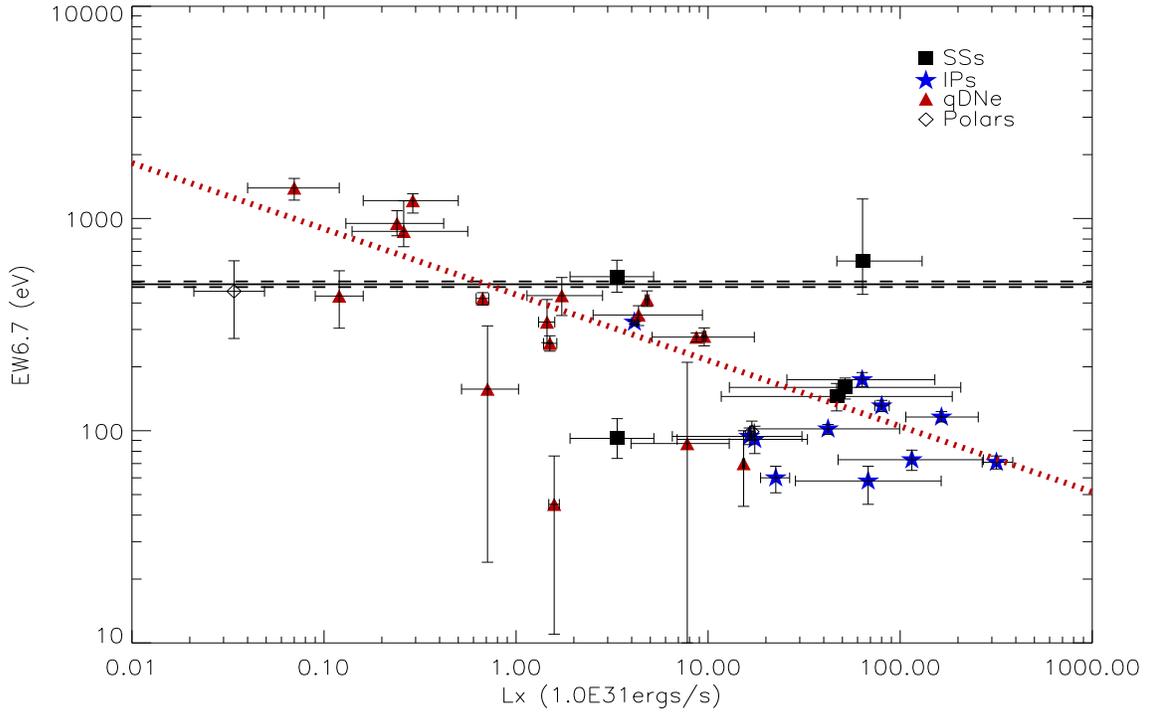}
   \caption{$EW_{6.7}$ vs. $L_{\rm 2-10keV}$  of our sample CVs. The solid and dashed black lines mark the $EW_{6.7}$ value and  error range of the GRXE. The red dotted line represents the best-fitted log-log linear relation for DNe only (see the text for details). 
}
   \label{Fig:07}
\end{figure}

\begin{figure}
\centering
\includegraphics[height=4in,width=6in]{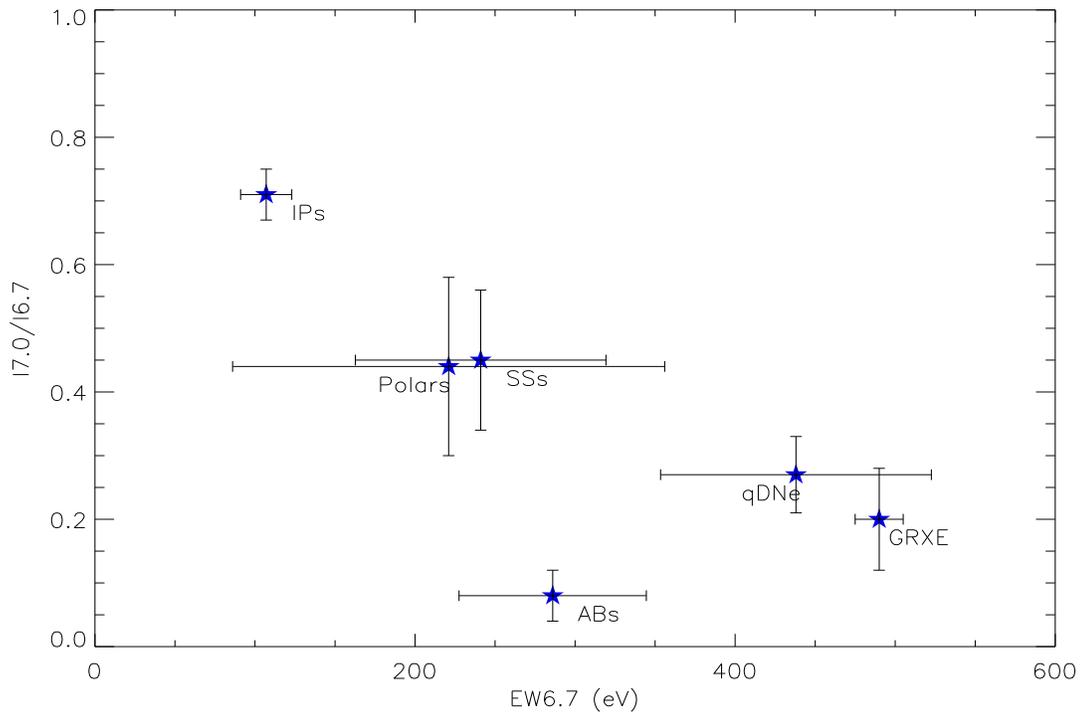}
   \caption{Mean $I_{7.0}/I_{6.7}$ vs. mean $EW_{6.7}$ of our sample classes, compared with the values of the GRXE. }
   \label{Fig:08}
\end{figure}

Fig~\ref{Fig:06} compares $I_{6.4}/I_{6.7}$ with $I_{7.0}/I_{6.7}$ of our sample sources. As
expected, there is a correlation between the two parameters, even for  two individual classes,  DNe and IPs. Therefore, $I_{6.4}/I_{6.7}$ is also a reasonably good tracer of the hot plasma temperature for these classes. SSs, on the other hand, can have exceptionally stronger 6.4-keV line emission; three of them have $I_{6.4}/I_{6.7} \gtrsim 2.5$, which can be naturally explained by the presence of cool (flureschncing) material surrounding such systems.

Theoretically, one may also expect an anti-correlation of $EW_{6.7}$ with $L_{\rm X}$ or $I_{7.0}/I_{6.7}$. For a same accretion rate, the deeper the gravitational potential is (as traced by $EW_{6.7}$ and $I_{7.0}/I_{6.7}$ ), the higher  $L_{\rm X}$ should be. Indeed,
Fig~\ref{Fig:07} shows these expected correlations, most apparent when data points with large errors are discounted. All IPs, except for one, have $L_{\rm X}>10^{32}{\rm~erg~s^{-1}}$, while almost all DNe have $L_{\rm X} < 10^{32}{\rm~erg~s^{-1}}$. The correlations are particularly convincing for DNe as a class with the Spearman's rank order correlation coefficient $r_s=0.65\pm0.17$. These correlations are not expected to be tight, because $L_{\rm X}$ also depends on the accretion rate, which could vary from one source to another of similar WD masses. Following the procedure described in~\citet{li13}, we characterize the $EW_{6.7}$-$L_{\rm X}$ correlation of DNe, using the best-fit log-log linear relation, $EW_{6.7}=(438\pm95 {\rm~eV}) (L/10^{31}~{\rm~ergs~s^{-1} })^{(-0.31\pm0.15)}$ (Fig.~\ref{Fig:07}). The root mean square of the data around the relation is $\approx 0.33\pm0.11$ dex. 

Fig~\ref{Fig:08} shows that the GRXE (especially its $EW_{\rm 6.7}$ value) 
cannot be explained by any mixture of mCVs, SSs, and ABs, if our sample sources 
are representative of these individual classes. mCVs have higher 
$T_{\rm b}$, lower $EW_{\rm 6.7}$ and higher $I_{7.0}/I_{6.7}$, compared to the GRXE.  
Similarly, ABs on average still have too low $EW_{\rm 6.7}$, but also too low 
$I_{7.0}/I_{6.7}$. In contrast, DNe have Fe lines properties consistent with those of 
the GRXE with in 1-$\sigma$. We discuss the implications of these 
results in the following section.

\section{Discussion}

The above spectral results are indicative as to the relative importance of
the source classes responsible for the GRXE. However, to quantify their respective
contributions, we need to be aware of the limitations of the results and
to know the flux distributions of the source classes. To do so, we first 
compare our results with those from previous studies as a consistency check 
and then discuss the limitations and their effects as well as the 
ways to move forward.

\subsection{Comparison to previous studies and common limitations}
Various studies of local CVs and ABs have been carried out previously.
The same \textit{Suzaku} data on the IPs and SSs in our sample have mostly 
been analyzed; the results are quantitatively consistent with ours, in terms of 
$T_{\rm b}$, as well as the Fe line equivalent widths (e.g., V1223 Sgr and V407 Cyg; 
\citealt{hay11,muk12}). Only 8 of the CVs and 4 of the ABs 
are analyzed for the first time; they are marked in Table 1. All of these
newly analyzed CVs are DNe, compared to eight studied previously \citep{ish09,iki13,muk09,byc10,sai12,neu14}). Some of our sample sources are included in existing \textit{ASCA} studies, the results of which typically have large uncertainties. The EW uncertainty is typically $\sim 40 $~eV in \textit{ASCA} measurements~\citep{ish99,ezu99}, compared to $\sim 10 $~eV in our \textit{Suzaku} ones; the latter tend to have substantially higher signal to noise ratios, as well as the improved spectral resolution, than the former. Our results are also consistent with previous \textit{Chandra HETG} measurements \citep[e.g., for SS Cyg and U Gem,][]{ran06,sch14}. Recently, \citet{eze15} compared the EWs of the 6.4~keV line of CVs to that of the GRXE based on a similar sample of SSs and mCVs, and conclude that SSs could account for the majority of the 6.4~keV line of GRXE. Our measurements of EWs of Fe lines are in general consistent with theirs. However, their work mostly focused on 6.4 keV lines, and did not include DNe, thus did not address the question of main contributors of 6.7/7.0-keV lines of GRXE as was done in this paper. Therefore, our results, broadly consistent with the existing ones, represent a step forward in improving both the sample 
statistics and the uniformity of the data analysis, focusing on the 
Fe-line diagnostics.

There are various limitations in our results. First, the sample size of our 
studied sources remains small, especially for AB, SS, and Polar classes. Although they 
are unlikely to be major contributors to the observed 6.7-keV line 
emission of the GRXE (e.g., Fig.~\ref{Fig:08}), their contributions
to the 6.4-keV line (probably chiefly due to SSs) and to the soft X-ray 
continuum (from ABs) could be substantial, depending on their uncertain 
overall populations in the Galaxy. We thus focus our discussion on the relative roles 
of IPs and DNe, which are probably the two most important classes for explaining
the 6.7 and 7.0~keV lines of the GRXE.

Second, the energy range effectively covered by the \textit{Suzaku} data 
is limited. The energy coverage of XIS is mostly below 10 keV which brings uncertainties to temperature measurements. The hard X-ray detector (HXD) spectra, on the other hand, usually have sufficiently high signal-to-noise bins only up to $\sim 40$ keV, with typically only 2-3 bins above 20 keV. When the temperature
of a plasma reaches above $\sim 20-40$~keV, the line diagnostics are also of 
little use, because atoms are almost fully ionized. Consequently, in such
a case, the temperature cannot be well constrained. This limitation affects our
ability to investigate the very hard part of the GRXE. Accordingly, we limit
our discussion on the nature of the GRXE in the 2-10~keV band.

Third, our sample is certainly very incomplete and biased toward
relatively bright sources. The sampling is probably reasonable
for IPs and perhaps SSs, which tend to have high luminosities, but is
clearly inadequate for DNe and ABs. Only five sources in eight observations in our sample are less 
luminous than $10^{31}{\rm~erg~s^{-1}}$ (Fig. 7). Therefore, the line or 
temperature measurements based on the sample may not be representative.
In such a case (e.g., for DNe), the luminosity-dependence of the 
measurements must be accounted for. 

\subsection{Integrated DNe contribution to the GRXE}
To quantify the contribution of DNe to the GRXE, we need to know their LF.
Indeed, the LFs of local CVs and ABs have been studied; the statistics of 
DNe, which are mostly below several $10^{31}{\rm~erg~s^{-1}}$, remains too poor to be analyzed separately~\citep{saz06}. 
\citet{byc10,pre12,reis13} suggest the presence of a faint population of DNe below  $10^{31}{\rm~erg~s^{-1}}$ and a LF with $\frac{dN}{dL} \propto L^{-1.64}$; however, their sample sizes are also not large enough to put tight constraints on LF parameters, and the slope of the LF is inconsistent with that of the GRXE sources ($\frac{dN}{dL} \propto L^{-2.5}$, tighter constraints could be put to the LF of local DNe though, see discussion below). Here we use the well-constrained LF of 
X-ray sources directly observed in the limited window~\citep{rev09}: 
$\frac{dN}{dL} \propto L^{-2.5}$. This LF extends to the 
luminosity limit of $\sim 10^{30}{\rm~erg~s^{-1}}$. Between this limit
and $L \sim 10^{31}{\rm~erg~s^{-1}}$, the sources should mostly be 
DNe. Sources below $10^{30}~{\rm~ergs~s^{{-1}} }$, as suggested by \citet{byc10,pre12,reis13}, could be populous and contribute a negligible part of the GRXE. However, they are unlikely to dominate GRXE, because sources above $10^{30}~{\rm~ergs~s^{{-1}} }$ contribute more than 80\% of GRXE\citep{rev09}, which leaves only $20\%$ for fainter sources. Nevertheless, detailed investigation on their Fe line properties are necessary to confirm their relative importance. We will thus concentrate on sources between $10^{30-32}~{\rm~ergs~s^{{-1}} }$. Our objective here is 
to demonstrate how the luminosity-dependence affects the measurements of $EW_{\rm 6.7}$.

We calculate the LF-accumulated $EW_{6.7,A}$ as
\begin{equation}
  EW_{\rm 6.7}= \frac{\int_{L_{min}}^{L_{max}} EW_{\rm 6.7} L \frac{dN}{dL} \,dL}{L_{A}},
\end{equation}
where the continuum at 6.7~keV is assumed to be scaled with 
$L$ and 
\begin{equation}
  L_{A}= \int \frac{dN}{dL} L\,dL.
\end{equation}
We account for the $EW_{6.7}$ dependence on $L$ using the best-fit relation obtained
in \S~3. If we take the upper and lower luminosity limits of the integration as $10^{32}~{\rm~ergs~s^{{-1}} }$ and $10^{30}~{\rm~ergs~s^{{-1}} }$, to match the luminosity range of the observed LF and DNe,  
the resulting $EW_{6.7}$ is $540_{-88}^{+79}$eV (or $448_{-65}^{+60}$eV if the upper luminosity limit taken to $10^{34}~{\rm~ergs~s^{{-1}} }$), which is consistent with, or slightly greater than the measured GRXE value. 
An excess of the DN $EW_{6.7}$ value
is indeed needed to balance the contributions from mCVs and
ABs, which have substantially lower $EW_{6.7}$ values ($EW_{6.7}$ below $300$~eV, see Table 3.). 

Conversely, we can use the observed $EW_{6.7}=490\pm15$~eV value of the GRXE to 
place constraints on the LF slope of local DNe by integrating EQ. 4, if we assume CVs in both environments share similar $EW_{6.7}$ values. To reach a $EW_{6.7}=490\pm15$~eV, the LF has to be steeper than $-2.25$ over the $10^{30}- 10^{34}~{\rm~ergs~s^{{-1}} }$ range. Of course, accounting for contributions from other source populations would require a steeper LF and/or an even lower luminosity limit of DNe. Therefore, our results are consistent with the presence of a large population of local DNe with individual 
luminosities below $10^{31}~{\rm~ergs~s^{{-1}} }$, (more than ten times more numerous
than those above $10^{31}~{\rm~ergs~s^{{-1}} }$; \citet{byc10,pre12,reis13}), but requires a steeper LF which needs to be confirmed by further observations. 

\subsubsection{Contributions from other sources}

We have focused on the Fe emission lines and their constraints on the nature
of the sources responsible for the GRXE. As shown in Fig.~1, the lines are good
diagnostics of the shock temperature (for the cooling flow model) up to about 20-40 keV.
The line emission saturates at higher temperatures (e.g., $EW_{6.7}$ predicted by cooling flow model hardly changes above 40 keV, and $EW_{6.7}$ by {\sl apec} model drops below 10 eV, which is hard to be measured; meanwhile, $I_{7.0}/I_{6.7}$ of cooling flow model becomes flatter and that of {\sl apec} model starts dropping above 40 keV). This saturation is 
the reason that mCVs cannot be the major contributor to the strong 
line emission observed in the GRXE. However, its hard X-ray continuum spectrum 
is likely dominated by mCVs. 

As shown in Fig.~\ref{Fig:03}, mCVs (especially IPs) typically have 
$T_{\rm b} > 15 $~keV, higher than the characteristic temperature of the GRXE 
in the 2-50 keV band\citep{yua12}. Because $T_{\rm b}$ is lower than $T_{\rm shock}$ in the cooling-flow spectral model of mCVs,
their mean continuum shape should be harder than that indicated by the one-temperature
characterization. Such a hard continuum contribution from mCVs can naturally be 
balanced by the soft mean spectra of DNe and ABs, which all have characteristic
temperatures lower than 15~keV. 

In short, while a quantitative decomposition of the GRXE into the contributions
from specific X-ray source classes is still beyond the scope of the present study,
it is clear that their relative importances are different 
for the Fe lines and for the hard spectral continuum. 

\section{Summary}

We have systematically analyzed of the  {\sl Suzaku}  spectra of 45 local CVs and ABs to explore the significance of their contributions to the GRXE. This sample consists of 
6 SSs, 16 IPs, 3 polars, 16 DNe and 4 ABs. The data for 12 of these sources 
are analyzed for the first time. Our main results and conclusions are as follows:

\begin{itemize} 
\item
Our measurements are focused on the Fe 6.4, 6.7, 7.0-keV lines, which are used as 
temperature diagnostics of the X-ray-emitting optically-thin thermal plasma. 
We find that the mean EW of the 6.7-keV line decreases from $\sim 438 (286)$ eV for DNe (ABs) to $\sim 107$ eV for IPs. In contrast, the line flux ratio $I_{7.0}/I_{6.7}$ increases from $\sim$ 0.27 for DNe to $\sim$ 0.71 for IPs. Such trends are well consistent with the expected higher temperature plasma in IPs due to the greater in-falling velocities of accreted matter than in DNe of similar WD masses. We also find that $I_{6.4}/I_{6.7}$ is strongly correlated with $I_{7.0}/I_{6.7}$ (except for the SSs), consistent with the expectation that the 6.4-keV line also traces the Fe-ionizing photon fluxes and 
hence the plasma temperature.

\item All sources classes considered here, except for DNe, show 
average Fe-line diagnostics ($EW_{\rm 6.7}$ and $I_{7.0}/I_{6.7}$) significantly different 
from those observed in the GRXE. Thus, DNe, though relatively faint  
($10^{30-32}{\rm~erg~s^{-1}}$ in our sampled range),
 are the most likely class accounting 
for the bulk of the 6.7-keV and 7.0-keV line intensities observed in the GRXE. 
In particular, we find a strong correlation between $EW_{6.7}$  and $L$, which can 
be characterized by the relation 
$EW_{6.7}=(438\pm95 {\rm~eV}) (L/10^{31}~{\rm~ergs~s^{-1} })^{(-0.31\pm0.15)}$.
This correlation indicates that the bulk of the emission arises from faint DNe.

\item Without a major contribution from DNe, no combination of other source classes
seems to be able to explain the observed 6.7-keV and 7.0-keV line intensities 
of the GRXE. These classes, however, can be major contributors to other parts of
the GRXE. In particular, IPs can dominate in the harder band. 
IPs are in general brighter than $10^{32}{\rm~erg~s^{-1}}$. Therefore, relatively 
bright sources observed in the GRXE tend to be mCVs. 
Conversely, ABs may be significant in contributing to the soft X-ray continuum.
\end{itemize} 

These results demonstrate the diagnostic power of the Fe emission lines. In comparison with the overall X-ray spectral shape, the EWs of the lines or their ratios provide more direct probes of the X-ray-emitting plasma in the intermediate temperature range
of $\sim 1 - 20$~keV and are less sensitive to complications such as the metallicity of the plasma, as well as the reflection and absorption of X-rays. The observed lines of the GRXE can therefore be used as the fundamental constraints and limits on 
the nature and population of the underlying responsible source classes.

\acknowledgements

The authors thank the anonymous referee for constructive comments that helped improve this paper. WQD appreciate valuable discussions with K. Mukai and T. Yuasa during the course of this work. This work is partly supported by National Science Foundation of China through grants NSFC-11303015 and NSFC-11133001.

\begin{table}[h]
\caption[]{Basic information of sampled CVs \& ABs.}
\label{tbl-01}
\tiny
  \begin{center}\begin{tabular}{ccccccc}
  \hline\hline
Source & Class & Obs-ID & D & $F_{\rm 2-10 }$ &$L$ & References \\
\hline
       &       &        & (kpc) & $10^{-12}{\rm~erg~s^{-1}~cm^{-2}}$ & $10^{31}{\rm~erg~s^{-1}}$& \\
\hline
CH Cyg$_{1}$  &SS& 400016020 & $0.268_{-0.066}^{+0.066}$ & 3.89 & $3.35_{- 1.44}^{+1.85}$ &\citet{per97}\\
CH Cyg$_{2}$  &SS& 400016030 & $0.268_{-0.066}^{+0.066}$ & 3.91&$ 3.36_{-  1.45 }^{+1.86} $&\citet{per97}\\
RS Oph   &SS& 406033010& $1.4 _{-0.2 }^{+0.6}$  & 2.71 &  $63.7 _{-16.9}^{+ 66.3}$&\citet{bar08}\\
RT Cru  &SS& 402040010& $- $ & 30.0 &  $-$&$-$\\
SS73-17$_{1}$& SS&401055010 & $0.5_{- 0.25}^{+ 0.5}$  &  17.2 & $  51.7 _{-38.8}^{+ 155}$&\citet{per97}\\
SS73-17$_{2}$& SS& 403043010& $0.5_{- 0.25}^{+ 0.5 }$ &  15.6 & $ 46.7 _{-35.0 }^{+140}$&\citet{per97}\\
T Crb& SS& 401043010& $- $ &  36.1  & $-$&$-$\\
V407 Cyg& SS& 905001010& $- $ &   6.99  & $ -$&$-$\\
AO Psc& IP& 404033010& $0.33_{- 0.12}^{+ 0.18} $&   4.86& $  63.4_{- 37.7 }^{+88.1}$&\citet{pre14}\\
BG Cmi& IP& 404029010& $-$ &   23.0 &  $ -$ &$-$\\
EX Hya& IP&402001010& $0.065_{- 0.001}^{+ 0.001 }$ &  81.7  & $ 4.13_{- 0.13}^{+ 0.13}$&\citet{god12}\\
FO Aqr& IP& 404032010& $0.45_{- 0.16 }^{+0.24}$  &  47.6 & $  115_{- 67.5}^{+ 156}$&\citet{pre14}\\
IGR J17195-4100 & IP& 403028010&$- $ &  34.3  & $ -$&$-$\\
IGR J17303-0601 & IP& 403026010&$-$  &  17.9  & $ -$&$-$\\
MU Cam & IP & 403004010&$- $ &  9.65 &$   - $&$-$\\
NY Lup & IP & 401037010&$0.68_{- 0.13}^{+ 0.17} $ &  29.6  &$  164 _{-57}^{+ 91}$&\citet{dem06}\\
PQ Gem & IP & 404030010&$0.51 _{-0.18}^{+ 0.28 }$ &  21.9 &  $ 68.1_{- 39.6 }^{+95.4 }$&\citet{pre14}\\
1RXS J213344.1+51072&IP & 401038010& $- $&  20.1  & $ -$&$-$\\
TV Col & IP & 403023010& $0.368_{- 0.015 }^{+0.017 }$ &  49.5 & $  80.1_{- 6.40}^{+ 7.57}$&\citet{ozd15}\\
TX Col & IP & 404031010& $-$ &   10.8 &  $ -$ &$-$\\
V1223 Sgr & IP & 402002010& $0.527_{- 0.043}^{+ 0.054 }$ &  95.6 &  $ 317_{- 49.7 }^{+68.4}$&\citet{ozd15}\\
V2400 Oph& IP & 403021010& $0.28_{- 0.10}^{+ 0.15 }$ &  44.9&   $42.1_{- 24.7 }^{+57.2}$&\citet{bar06}\\
V709 Cas& IP &403025010& $0.230_{- 0.020}^{+ 0.020}$  &  35.5 & $  22.5_{- 3.74 }^{+4.08}$&\citet{bar06}\\
XY Ari$_{1}$ & IP &500015010 & $0.27 _{-0.1}^{+ 0.1 }$ &  18.8 &  $ 16.4 _{-9.89}^{+ 14.4}$&\citet{pre14}\\
XY Ari$_{2}$ & IP & 506026010& $0.27_{- 0.1}^{+ 0.1}$  &  20.0 &  $ 17.5_{- 10.6}^{+ 15.4}$&\citet{pre14}\\
SWIFT J2319.4+2619 &Po &408030010 & $ - $ &  3.08 &  $ -$&$-$\\
AM Her & Po & 403007010& $0.091_{-0.015}^{+ 0.018} $ &  0.350  & $ 0.034_{- 0.013}^{+ 0.015}$ &\citet{gan95}\\
V1432 Aql & Po & 403027010& $0.23  $ &  26.7 &  $ 16.9\pm0.3 $&\citet{wat95}\\
BF Eri$^{a}$ & DN & 407045010& $0.7_{- 0.2 }^{+0.2 }$ &  1.33  &  $7.79_{- 3.82}^{+ 5.09}$&\citet{neu08}\\
BV Cen$^{a}$ & DN &407047010 & $0.238 $ &  14.1  & $ 9.54\pm0.2$&\citet{god12}\\
BZ UMa & DN & 402046010& $0.228_{- 0.043}^{+ 0.063 }$ &   2.79 &   $1.73_{- 0.59 }^{+1.09}$&\citet{tho08}\\
EK TrA$^{a}$ & DN &407044010 & $0.18  $ &  4.09 &  $1.58\pm0.1 $ &\citet{gan97}\\
FL Psc$^{a}$ & DN & 403039010& $- $ &  0.396 &  $ -  $&$-$\\
FS Aur & DN & 408041010& $- $ &  2.53  & $- $&$-$\\
GK Per$^{a}$ & DN & 403081010&$ 0.420  $ &  7.26 &  $ 15.3\pm0.2$&\citet{pre14}\\
KT Per & DN & 403041010& $0.145 _{-0.021}^{+ 0.030}$ &  2.82  & $ 0.708 _{-0.191}^{+ 0.321}$&\citet{tho08}\\
SS Aur & DN & 402045010& $0.201_{- 0.010 }^{+0.010 }$ &  2.99 &  $ 1.45_{- 0.14}^{+ 0.14}$&\citet{ozd15}\\
SS Cyg & DN & 400007010& $0.166 $ &  14.6 &  $ 4.81\pm0.04 $&\citet{ozd15}\\
U Gem$_{1}$ & DN & 407034010& $0.1_{- 0.004 }^{+0.004 }$ &  12.5 &  $ 1.50 _{-0.11 }^{+0.13}$&\citet{ozd15}\\
U Gem$_{2}$ & DN & 407035010& $0.1_{- 0.004 }^{+0.004 }$ &  5.56 &  $0.665 _{-0.052}^{+ 0.054}$&\citet{ozd15}\\
V1159 Ori$^{a}$& DN & 408029010& $- $ &  1.00 & $ -$&$-$\\
V893 Sco& DN & 401041010& $0.135_{- 0.032}^{+ 0.063 }$ &  19.9 &   $4.34_{- 1.82}^{+ 5.00}$&\citet{ozd15}\\
VW Hyi$_{1}$$^{a}$ & DN & 406009010& $0.064 _{-0.017}^{+ 0.020}$ &   1.44 &   $0.071_{- 0.032}^{+ 0.051}$&\citet{pre12}\\
VW Hyi$_{2}$$^{a}$ & DN & 406009020& $0.064_{- 0.017 }^{+0.020} $ &  5.86 &  $0.287_{- 0.133}^{+ 0.207}$&\citet{pre12}\\
VW Hyi$_{3}$$^{a}$ & DN & 406009030& $0.064 _{-0.017}^{+ 0.020 }$ &  4.96 &   $0.243_{- 0.112}^{+ 0.175}$&\citet{pre12}\\
VW Hyi$_{4}$$^{a}$ & DN & 406009040& $0.064_{- 0.017}^{+ 0.020}$ &   5.20  & $ 0.255_{- 0.117 }^{+0.301}$&\citet{pre12}\\
VY Aqr$^{a}$ & DN & 402043010& $0.097_{- 0.012}^{+ 0.015} $ &  1.08  & $ 0.122_{-0.28}^{+ 0.41}$&\citet{ozd15}\\
Z Cam & DN & 404022010& $0.163_{- 0.038}^{+ 0.068}$  &  27.4 &  $ 8.69_{- 3.58}^{+8.71}$&\citet{ozd15}\\
GT Mus$^{a}$ & AB & 402095010& $-$&$7.13$&$-$&$-$\\
II Peg$^{a}$ & AB & 407038010&$-$&$2.30$&$-$ &$-$\\
$\sigma$ Gem$^{a}$ & AB & 402033010&$-$&$1.01$&$-$&$-$\\
UX Ari$^{a}$ & AB & 404008010&$-$&$1.51$&$-$ &$-$\\
\hline
\end{tabular}
\end{center}
Note: $^{a}$Sources firstly analyzed in this paper.
\end{table}

\begin{table}[h]
\caption[]{Fitting results for individual sources}
\label{tbl-02}
\tiny
  \begin{center}\begin{tabular}{ccccccccccc}
  \hline\hline
Source & Class &$T_{\rm b}$ &$EW_{\rm 6.4}$ & $I_{6.4}/I_{6.7}$ & $E_{\rm 6.7}$ & $I_{6.7}$ & $EW_{\rm 6.7}$ &$EW_{\rm 7.0}$ & $I_{7.0}/I_{6.7}$ & $\chi^2/ \rm {d.o.f.}$ \\
\hline
 &  &(keV) &eV &  & (eV) &  & (eV) &(eV) &  & \\
\hline
CH Cyg$_{1}$  &SS& $200_{-54.6}^{+0.0}$ & $390_{-85}^{+88}$ & $2.91_{-0.42}^{+0.64}$ &  $6.56_{-0.03}^{+0.04}$ & $16.9_{-2.78}^{+2.31}$ & $532_{-83}^{+104}$ & $76_{-40}^{+40}$ & $0.56_{-0.11}^{+0.25}$ &1.52/191  \\
CH Cyg$_{2}$  &SS& $8.38_{-1.83}^{+2.97}$ & $839_{-68}^{+90}$ & $2.82_{-0.41}^{+0.53}$ &  $6.62_{-0.02}^{+0.02}$ & $30.4_{-4.12}^{+4.26}$ & $92_{-18}^{+22}$ &  $117_{-35}^{+32}$ & $0.33_{-0.09}^{+0.10}$ &1.05/172  \\
RS Oph   &SS& $3.46_{-2.20}^{+40.8}$ & $82_{-82}^{+193}$ & $2.72_{-1.07}^{+2.57}$& $6.71_{-0.09}^{+0.09}$ & $5.28_{-2.60}^{+2.51}$& $630_{-191}^{+606}$ &  $165_{-62}^{+244}$ &$0.22_{-0.22}^{+0.52}$ &0.91/49\\
RT Cru  &SS& $53.1_{-5.17}^{+8.61}$ & $200_{-13}^{+13}$ &$1.65_{-0.15}^{+0.18}$& $6.673_{-0.01}^{+0.01}$&$48.0_{-4.97}^{+4.76} $ & $88_{-9}^{+8}$ &  $70_{-11}^{+9}$ & $0.57_{-0.09}^{+0.11}$ &1.06/942\\
SS73-17$_{1}$& SS& $8.33_{-1.74}^{+2.73}$ & $176_{-20}^{+20}$& $1.06_{-0.15}^{+0.16}$ & $6.664_{-0.01}^{+0.01}$& $79.9_{-8.47}^{+11.0}$ & $160_{-19}^{+17}$ &  $109_{-22}^{+19}$ & $0.51_{-0.10}^{+0.11}$ & 1.02/423\\
SS73-17$_{2}$& SS& $13.25_{-3.04}^{+8.34}$ & $291_{-33}^{+35}$& $0.96_{-0.15}^{+0.20}$ & $6.69_{-0.02}^{+0.01}$& $50.6_{-9.13}^{+9.41}$ & $145_{-21}^{+22}$ &  $133_{-26}^{+36}$ & $0.54_{-0.11}^{+0.13}$ & 0.88/219\\
T Crb& SS& $34.6_{-2.72}^{+3.08}$ & $161_{-10}^{+10}$& $1.48_{-0.14}^{+0.17}$ & $6.70_{-0.01}^{+0.01}$& $52.7_{-5.15}^{+5.15}$ & $87_{-9}^{+8}$ &  $123_{-12}^{+13}$ & $0.83_{-0.10}^{+0.12}$ & 0.97/1229\\
V407 Cyg& SS& $3.96_{-0.53}^{+0.69}$ & $104_{-104}^{+0}$ & $0.05_{-0.05}^{+0.09}$ & $6.67_{-0.01}^{+0.01}$& $18.1_{-5.15}^{+5.15}$ & $418_{-51}^{+58}$ &  $0_{-0}^{+10}$ & $0.03_{-0.03}^{+0.08}$ & 1.10/116\\
AO Psc& IP& $19.7_{-2.49}^{+3.26}$ & $158_{-15}^{+15}$& $0.68_{-0.06}^{+0.07}$ & $6.67_{-0.01}^{+0.01}$& $114.8_{-7.52}^{+7.16}$ & $174_{-13}^{+14}$ &  $100_{-10}^{+10} $ & $0.56_{-0.06}^{+0.06}$ & 0.969/1262\\
BG Cmi& IP& $42.6_{-5.34}^{+6.55}$ & $102_{-16}^{+15}$& $1.18_{-0.19}^{+0.30}$ & $6.65_{-0.02}^{+0.02}$& $23.9_{-3.63}^{+3.55}$ & $81_{-12}^{+14}$ &  $54_{-13}^{+14} $ & $0.57_{-0.14}^{+0.18}$ & 1.04/934\\
EX Hya lines& IP& $9.41_{-0.38}^{+0.40}$ & $32_{-3}^{+3}$& $0.13_{-0.01}^{+0.01}$ & $6.67_{-0.01}^{+0.01}$& $310.6_{-5.39}^{+5.39}$ & $325_{-7}^{+6}$ &  $110_{-4}^{+4} $ & $0.39_{-0.01}^{+0.02}$ & 1.03/3100\\
FO Aqr& IP& $19.1_{-2.28}^{+2.88}$ & $133_{-9}^{+9}$& $1.75_{-0.19}^{+0.24}$ & $6.66_{-0.01}^{+0.02}$& $60.3_{-6.24}^{+6.21}$ & $73_{-8}^{+8}$ &  $58_{-11}^{+10} $ & $0.58_{-0.12}^{+0.12}$ & 0.94/1841\\
IGR J17195-4100 & IP& $30.5_{-6.06}^{+8.94}$ & $128_{-11}^{+10}$& $1.26_{-0.17}^{+0.20}$ & $6.68_{-0.01}^{+0.01}$& $45.0_{-4.94}^{+5.06}$ & $91_{-10}^{+10}$ &  $32_{-3}^{+2} $ & $0.87_{-0.14}^{+0.26}$ & 0.98/1131\\
IGR J17303-0601 & IP & $63.6_{-24.0}^{+64.8}$ & $88_{-24}^{+35}$ & $1.50_{-0.71}^{+0.99}$ & $6.54_{-0.08}^{+0.05}$ & $14.8_{-5.51}^{+4.07}$ & $59_{-18}^{+15}$ &  $62_{-16}^{+21} $ & $0.92_{-0.30}^{+0.66}$ & 0.91/721\\
MU Cam & IP & $15.8_{-3.04}^{+4.07}$ & $139_{-16}^{+17}$ & $1.21_{-0.19}^{+0.25}$ & $6.67_{-0.01}^{+0.01}$ & $23.0_{-3.07}^{+3.07}$ & $101_{-13}^{+14}$ &  $134_{-17}^{+18} $ & $0.94_{-0.16}^{+0.21}$ & 0.93/631\\
NY Lup & IP & $43.5_{-7.53}^{+11.76}$ & $156_{-7}^{+7}$ & $1.17_{-0.07}^{+0.12}$ & $6.67_{-0.01}^{+0.01}$ & $54.8_{-2.89}^{+2.89}$ & $116_{-6}^{+7}$ &  $120_{-7}^{+7} $ & $0.81_{-0.07}^{+0.08}$ & 0.99/2319\\
PQ Gem & IP & $32.6_{-7.71}^{+12.7}$ & $128_{-14}^{+14}$ & $1.93_{-0.36}^{+0.57}$ & $6.65_{-0.02}^{+0.02}$ & $17.7_{-3.49}^{+3.47}$ & $58_{-13}^{+10}$ &  $62_{-13}^{+12} $ & $0.77_{-0.22}^{+0.30}$ & 1.05/938\\
1RXS J213344.1+51072 & IP & $65.9_{-13.3}^{+18.3}$ & $172_{-12}^{+10}$ & $1.83_{-0.19}^{+0.24}$ & $6.69_{-0.02}^{+0.01}$ & $24.9_{-2.56}^{+2.66}$ & $79_{-9}^{+8}$ &  $91_{-10}^{+11} $ & $0.88_{-0.07}^{+0.14}$ & 1.07/1429\\
TV Col & IP & $40.5_{-8.28}^{+12.9}$ & $120_{-11}^{+9}$ & $0.86_{-0.08}^{+0.06}$ & $6.67_{-0.01}^{+0.01}$ & $92.1_{-6.21}^{+6.40}$ & $131_{-8}^{+8}$ &  $104_{-11}^{+10} $ & $0.67_{-0.04}^{+0.08}$ & 1.14/1403\\
TX Col & IP & $26.9_{-7.19}^{+6.10}$ & $88_{-15}^{+15}$ & $0.71_{-0.15}^{+0.16}$ & $6.68_{-0.01}^{+0.01}$ & $18.5_{-2.45}^{+2.37}$ & $121_{-19}^{+16}$ &  $94_{-17}^{+20} $ & $0.69_{-0.14}^{+0.18}$ & 0.92/626\\
V1223 Sgr & IP & $26.6_{-3.47}^{+4.40}$ & $97_{-5}^{+6}$ & $1.29_{-0.09}^{+1.11}$ & $6.67_{-0.01}^{+0.01}$ & $110.5_{-7.02}^{+7.02}$ & $71_{-5}^{+5}$ &  $70_{-7}^{+6} $ & $0.80_{-0.08}^{+0.08}$ & 1.04/3000\\
V2400 Oph& IP & $22.8_{-1.78}^{+2.01}$ & $140_{-6}^{+6}$ & $1.24_{-0.06}^{+0.07}$ & $6.67_{-0.01}^{+0.01}$ & $65.8_{-2.96}^{+2.96}$ & $102_{-6}^{+5}$ &  $93_{-5}^{+7} $ & $0.73_{-0.05}^{+0.05}$ & 1.05/2992\\
V709 Cas& IP & $47.3_{-5.07}^{+6.50}$ & $131_{-10}^{+11}$ & $1.98_{-0.26}^{+0.39}$ & $6.67_{-0.01}^{+0.02}$ & $29.5_{-4.18}^{+4.30}$ & $60_{-9}^{+8}$ &  $69_{-11}^{+11} $ & $0.92_{-0.18}^{+0.23}$ & 1.02/1397\\
XY Ari$_{1}$ & IP & $39.6_{-8.21}^{+13.1}$ & $70_{-8}^{+8}$ & $0.76_{-0.10}^{+0.11}$ & $6.67_{-0.01}^{+0.01}$ & $24.6_{-2.23}^{+2.29}$ & $94_{-8}^{+9}$ &  $57_{-9}^{+8} $ & $0.54_{-0.09}^{+0.10}$ & 0.97/1454\\
XY Ari$_{2}$ & IP & $31.6_{-8.46}^{+13.2}$ & $80_{-12}^{+16}$ & $0.84_{-0.11}^{+0.12}$ & $6.66_{-0.01}^{+0.01}$ & $24.1_{-1.83}^{+1.83}$ & $91_{-13}^{+14}$ &  $65_{-9}^{+11} $ & $0.62_{-0.10}^{+0.11}$ & 0.90/1823\\
SWIFT J2319.4+2619 & Po & $9.02_{-2.28}^{+4.80}$ & $90_{-32}^{+52}$ & $0.46_{-0.19}^{+0.23}$ & $6.65_{-0.03}^{+0.03}$ & $8.34_{-1.72}^{+1.57}$ & $208_{-44}^{+80}$ &  $119_{-35}^{+46} $ & $0.53_{-0.20}^{+0.26}$ & 0.89/184\\
AM Her & Po & $8.42_{-2.09}^{+3.08}$ & $0.0_{-0.0}^{+0.01}$ & $0.0_{-0.0}^{+0.09}$ & $6.63_{-0.04}^{+0.04}$ & $1.35_{-0.48}^{+0.48}$ &  $454_{-182}^{+178}$ & $43_{-43}^{+133} $ & $0.09_{-0.09}^{+0.37}$ & 1.11/112\\
V1432 Aql & Po & $16.7_{-2.28}^{+3.01}$ & $110_{-12}^{+10}$ & $1.08_{-0.15}^{+0.18}$ & $6.67_{-0.01}^{+0.01}$ & $46.9_{-5.07}^{+5.12}$ & $99_{-9}^{+12}$ &  $63_{-12}^{+13} $ & $0.51_{-0.11}^{+0.12}$ & 0.90/1052\\
BF Eri & DN & $3.56_{-1.20}^{+2.21}$ & $223_{-140}^{+660}$ & $0.0_{-0.0}^{+0.0}$ & $6.75_{-0.16}^{+0.16}$ & $0.69_{-0.69}^{+0.96}$ & $87_{-87}^{+123}$ &  $0.0_{-0.0}^{+0.0} $ & $0.0_{-0.0}^{+0.31}$ & 0.95/53\\
BV Cen & DN & $13.3_{-2.77}^{+4.82}$ & $68_{-16}^{+21}$ & $0.26_{-0.06}^{+0.07}$ & $6.67_{-0.01}^{+0.01}$ & $46.8_{-3.80}^{+3.66}$ & $279_{-28}^{+26}$ &  $151_{-17}^{+22} $ & $0.53_{-0.08}^{+0.09}$ & 0.92/498\\
BZ UMa & DN & $5.12_{-1.94}^{+5.23}$ & $76_{-40}^{+49}$ & $0.22_{-0.12}^{+0.15}$ & $6.67_{-0.02}^{+0.02}$ & $9.99_{-1.87}^{+1.77}$ & $433_{-84}^{+94}$ &  $153_{-53}^{+58} $ & $0.40_{-0.15}^{+0.17}$ & 0.87/74\\
EK TrA & DN & $38.8_{-14.7}^{+35.4}$ & $197_{-55}^{+60}$ & $0.11_{-0.07}^{+0.08}$ & $6.67_{-0.01}^{+0.01}$ & $20.0_{-2.49}^{+2.46}$ & $45_{-34}^{+31}$ &  $512_{-80}^{+135} $ & $0.16_{-0.08}^{+0.08}$ & 0.97/114\\
FL Psc & DN & $8.46_{-4.55}^{+34.9}$ & $76_{-76}^{+340}$ & $0.16_{-0.16}^{+0.61}$ & $6.69_{-0.01}^{+0.01}$ & $2.93_{-0.98}^{+0.98}$ & $920_{-511}^{+660}$ &  $46_{-46}^{+200} $ & $0.10_{-0.10}^{+0.57}$ & 0.78/51\\
FS Aur & DN & $15.9_{-6.08}^{+17.5}$ & $52_{-28}^{+35}$ & $0.32_{-0.18}^{+0.23}$ & $6.67_{-0.03}^{+0.03}$ & $5.16_{-1.13}^{+1.09}$ & $189_{-51}^{+67}$ &  $76_{-35}^{+47} $ & $0.40_{-0.19}^{+0.24}$ & 0.97/252\\
GK Per & DN & $13.6_{-3.96}^{+8.95}$ & $77_{-31}^{+35}$ & $1.06_{-0.45}^{+0.78}$ & $6.65_{-0.08}^{+0.08}$ & $6.47_{-2.42}^{+2.47}$ & $70_{-26}^{+30}$ &  $78_{-30}^{+33} $ & $0.94_{-0.45}^{+0.76}$ & 1.13/210\\
KT Per & DN & $10.3_{-3.82}^{+12.8}$ & $37_{-36}^{+33}$ & $0.16_{-0.16}^{+0.18}$ & $6.71_{-0.38}^{+0.20}$ & $4.73_{-4.34}^{+4.47}$ & $157_{-133}^{+155}$ &  $25_{-25}^{+57} $ & $0.10_{-0.10}^{+0.18}$ & 0.85/118\\
SS Aur & DN & $8.48_{-2.61}^{+5.34}$ & $47_{-29}^{+60}$ & $0.17_{-0.13}^{+0.15}$ & $6.66_{-0.02}^{+0.02}$ & $14.0_{-2.67}^{+2.66}$ & $325_{-65}^{+90}$ &  $169_{-49}^{+57}$ & $0.56_{-0.18}^{+0.23}$ & 0.89/84\\
SS Cyg & DN & $8.15_{-0.91}^{+1.23}$ & $67_{-10}^{+11}$ & $0.30_{-0.04}^{+0.05}$ & $6.67_{-0.01}^{+0.01}$ & $57.1_{-2.88}^{+2.89}$ & $415_{-30}^{+41}$ &  $48_{-11}^{+15}$ & $0.16_{-0.04}^{+0.04}$ & 1.12/773\\
U Gem$_{1}$ & DN & $16.5_{-3.31}^{+4.49}$ & $43_{-11}^{+16}$ & $0.19_{-0.05}^{+0.05}$ & $6.67_{-0.01}^{+0.01}$ & $19.4_{-1.30}^{+1.27}$ & $258_{-21}^{+22}$ &  $178_{-11}^{+11}$ & $0.68_{-0.08}^{+0.08}$ & 0.97/839\\
U Gem$_{2}$ & DN & $9.55_{-1.16}^{+1.54}$ & $94_{-14}^{+13}$ & $0.30_{-0.04}^{+0.06}$ & $6.67_{-0.01}^{+0.01}$ & $64.5_{-3.84}^{+3.22}$ & $420_{-29}^{+27}$ &  $100_{-14}^{+13}$ & $0.28_{-0.04}^{+0.04}$ & 1.06/690\\
V1159 Ori& DN & $6.97_{-1.83}^{+2.22}$ & $40_{-8}^{+48}$ & $0.15_{-0.15}^{+0.17}$ & $6.66_{-0.08}^{+0.07}$ & $2.87_{-0.42}^{+0.42}$ & $331_{-47}^{+105}$ &  $45_{-6}^{+68}$ & $0.14_{-0.14}^{+0.17}$ & 0.94/181\\
V893 Sco& DN & $9.36_{-1.21}^{+1.59}$ & $51_{-11}^{+14}$ & $0.19_{-0.05}^{+0.05}$ & $6.69_{-0.01}^{+0.01}$ & $75.3_{-4.98}^{+5.07}$ & $351_{-38}^{+37}$ &  $110_{-15}^{+19}$ & $0.37_{-0.06}^{+0.06}$ & 0.91/558\\
VW Hyi$_{1}$ & DN & $5.79_{-2.14}^{+4.71}$ & $0_{-0}^{+0.01}$ & $0.08_{-0.06}^{+0.07}$ & $6.65_{-0.01}^{+0.02}$ & $9.3_{-1.00}^{+1.01}$ & $1392_{-172}^{+152}$ &  $42_{-42}^{+58}$ & $0.07_{-0.07}^{+0.07}$ & 1.10/156\\
VW Hyi$_{2}$ & DN & $5.87_{-0.34}^{+0.41}$ & $0_{-0}^{+0.01}$ & $0.05_{-0.05}^{+0.06}$ & $6.66_{-0.01}^{+0.01}$ & $49.4_{-4.59}^{+4.59}$ & $1214_{-153}^{+95}$ &  $20_{-20}^{+34}$ & $0.04_{-0.05}^{+0.05}$ & 1.04/114\\
VW Hyi$_{3}$ & DN & $6.45_{-1.65}^{+2.81}$ & $0_{-0}^{+0.01}$ & $0.09_{-0.07}^{+0.07}$ & $6.68_{-0.01}^{+0.01}$ & $43.1_{-3.86}^{+3.86}$ & $948_{-119}^{+140}$ &  $92_{-35}^{+33}$ & $0.21_{-0.06}^{+0.08}$ & 1.16/131\\
VW Hyi$_{4}$ & DN & $4.81_{-1.14}^{+1.80}$ & $0_{-0}^{+0.01}$ & $0.05_{-0.05}^{+0.06}$ & $6.67_{-0.01}^{+0.01}$ & $38.5_{-4.02}^{+4.03}$ & $870_{-134}^{+341}$ &  $84_{-40}^{+33}$ & $0.19_{-0.08}^{+0.08}$ & 1.07/115\\
VY Aqr & DN & $12.0_{-6.20}^{+141}$ & $30_{-30}^{+63}$   & $0.12_{-0.12}^{+0.27}$   & $6.66_{-0.03}^{+0.03}$ & $5.36_{-1.61}^{+1.61}$ & $431_{-127}^{+136}$ &  $0_{-0}^{+0.01}$ & $0.0_{-0.0}^{+0.01}$ & 1.05/52 \\
Z Cam & DN & $12.5_{-1.43}^{+2.11}$ & $54_{-8}^{+9}$ & $0.23_{-0.04}^{+0.04}$ & $6.68_{-0.01}^{+0.01}$ & $108.5_{-5.13}^{+5.13}$ & $276_{-14}^{+13}$ &  $161_{-13}^{+12}$ & $0.58_{-0.05}^{+0.05}$ & 1.01/1098   \\
GT Mus & AB & $5.44_{-0.18}^{+0.19}$ & $15_{-5}^{+4}$ & $0.08_{-0.03}^{+0.03}$ & $6.67_{-0.01}^{+0.01}$ & $130.4_{-4.7}^{+4.6}$ & $259_{-11}^{+9}$ &  $18_{-6}^{+7}$ & $0.10_{-0.03}^{+0.03}$ & 1.03/1432   \\
II Peg & AB & $3.59_{-0.24}^{+0.25}$ & $0_{-0}^{+15}$ & $0.00_{-0.00}^{+0.08}$ & $6.67_{-0.01}^{+0.01}$ & $23.7_{-2.9}^{+2.9}$ & $217_{-35}^{+30}$ &  $0.1_{-0.1}^{+19}$ & $0.01_{-0.01}^{+0.09}$ & 1.08/276   \\
$\sigma$ Gem & AB & $2.66_{-0.21}^{+0.20}$ & $0_{-0}^{+11}$ & $0.0_{-0.0}^{+0.04}$ & $6.66_{-0.01}^{+0.01}$ & $14.3_{-1.11}^{+1.36}$ & $436_{-48}^{+45}$ &  $0_{-00}^{+17}$ & $0.0_{-0.0}^{+0.01}$ & 0.88/273   \\
UX Ari & AB & $2.61_{-0.18}^{+0.21}$ & $7_{-7}^{+15}$ & $0.04_{-0.04}^{+0.11}$ & $6.67_{-0.01}^{+0.01}$ & $11.8_{-1.11}^{+1.36}$ & $248_{-32}^{+35}$ &  $30_{-20}^{+30}$ & $0.16_{-0.12}^{+0.12}$ & 1.02/505   \\
\hline
\end{tabular}
\end{center}

Note: The fits are based on the absorbed zero-metallicity apec with the three gaussian model (\S~2). $I_{6.7}$ is in units of $10^{-14}{\rm ergs^{-1}cm^{-2}}$. The last column shows the reduced
$\chi^2$ and the degrees of freedom of each fit.
\end{table}

\begin{table}[h]
\caption[]{Mean values of key parameters for individual classes.}
\label{tbl-03}
  \begin{center}\begin{tabular}{cccccc}
  \hline\hline
Source Class & $T_{\rm b}$ &$EW_{\rm 6.4}$&$EW_{\rm 6.7}$  & $EW_{\rm 7.0}$ & $I_{7.0}/I_{6.7}$   \\
\hline
 & (keV)& (eV)&(eV)&(eV)&\\
\hline
SSs &    $27.2\pm20.8$ & $280\pm90.0 $ &$241\pm 78.3$ & $91\pm 20.1$ & $0.45\pm 0.11$\\
IPs &    $34.0\pm4.54$ & $115\pm9.12 $ &$107\pm 16.0$ & $80\pm 6.81$ & $0.71\pm 0.04$ \\
Polars & $11.4\pm2.84$ & $66.7\pm44.4 $ &$221\pm 135$ &  $74\pm 29.7$ & $0.44\pm 0.14$\\
DNe &    $10.7\pm2.04$ & $61.6\pm18.7 $ &$438\pm 84.6$ & $95\pm 18.6$ & $0.27\pm 0.06$\\
ABs &    $3.6\pm0.61$  & $5.50_{-5.50}^{+11.1} $ &$286\pm 58.5$ & $12.1\pm 7.36$ & $0.08\pm 0.04$\\
GRXE &   $15.1_{-0.7}^{+0.4}/6.64_{-0.42}^{+0.40}$ & $110_{-10}^{+9}$ &$490\pm 15$ & $110 \pm 10$ & $0.2\pm 0.08$\\
\hline
\end{tabular}
\end{center}

Note: The errors of the means are at the 1-$\sigma$ level. 
The GRXE values are from \citet{uch13} and \citet{yua12} are included for comparison. $T_{\rm b}$ of GRXE: 6.64 keV from \citet{uch13} of 2-10 keV range fitting; 15 keV from \citet{yua12} of 2-50 keV range fitting. See Discussion for details. 
\end{table}

\begin{thebibliography}{}

\bibitem[Anzolin et al.(2009)]{anz09}  Anzolin, G., de Martino, D., Falanga, M., et al., 2009, \aap, 501, 1047
\bibitem[Arnaud(1996)]{arn96}Arnaud K., 1996, in Jacoby G. H., Barnes J., eds, ASP Conf. Ser. Vol. 101, Astronomical Data Analysis Software and Systems V. Astron. Soc. Pac., San Francisco, p. 17
\bibitem[Barlow et al.(2006)]{bar06}Barlow, E. J., Knigge, C., Bird, A.J., et al., 2006, \mnras, 372, 224
\bibitem[Barry et al.(2008)]{bar08}	Barry, R. K., Mukai, K., Sokoloski, J.L., et al., 2008, ASPC, 401, 52
\bibitem[Baskill et al.(2005)]{bas05}Baskill, D. S., Wheatley, P. J., \& Osborne, J. P., 2005, \mnras, 357, 626
\bibitem[Berdyugina et al.(1998)]{ber98}Berdyugina, S.V., Jankov, S., Ilyin, I., et al., 1998, \aap, 334, 863
\bibitem[Byckling et al.(2010)]{byc10}Byckling, K., Mukai, K., Thorstensen, J. R., \& Osborne, J. P., 2010, \mnras, 408, 2298
\bibitem[de Martino et al.(2006)]{dem06}de Martino D., Bonnet-Bidaud, J.-M., Mouchet, M.,  et al., 2006, \aap, 449, 1151
\bibitem[Ebisawa et al.(2008)]{ebi08}Ebisawa, K., Yamauchi, S., Tanaka, Y., et al., 2008, PASJ, 60, S223
\bibitem[Eker et al.(2008)]{eke08}Eker, Z., Ak, N. F., Bilir, S., et al., 2008, \mnras, 389, 1722 
\bibitem[Eze(2015)]{eze15} Eze, R.N.C., New Astronomy, 36, 64 
\bibitem[Ezuka \& Ishida(1999)]{ezu99}Ezuka, H., \& Ishida, M., 1999, \apjs, 120, 277
\bibitem[Franciosini et al.(2001)]{fra01}Franciosini, E., Pallavicini, R., \& Tagliaferri, G., 2001, \aap, 375, 196	
\bibitem[Gansicke(1995)]{gan95}Gansicke, B., Beuermann, K., \& de Martino, D., 1995, \aap, 303, 127
\bibitem[Gansicke et al.(1997)]{gan97}Gansicke, B., Beuermann, K., \& Thomas, H., 1997, \mnras, 289, 388
\bibitem[Gansicke et al.(2009)]{gan09}Gansicke B.T., Dillon, M., Southworth, J., et al., 2009, \mnras, 397, 2170	
\bibitem[Godon et al(2012)]{god12}Godon, P., Sion, E. M., Levay, K., et al., 2012, \apjs, 203, 29
\bibitem[Gray (2008)]{gray08}Gray, D., 2008, The Observation and Analysis of Stellar Photospheres, Cambridge University Press, 3rd edition 
\bibitem[Hayashi et al.(2011)]{hay11}Hayashi, T., Ishida, M., Terada, Y., et al., 2011, PASJ, 63, 739
\bibitem[Hellier \& Mukai(2004)]{hel04}Hellier, C., Mukai, K., 2004, \mnras, 352, 1037
\bibitem[Hong et al.(2012)]{hon12}Hong, J., van den Berg, M., Grindlay, J. E.,  et al., 2012, \apj, 746, 165
\bibitem[Ikis et al.(2013)]{iki13}Ikis Gun, G, Karagul, A., \& Gok, F., 2013, New Astronomy, 25, 1
\bibitem[Ishida \& Ezuk(1999)]{ish99}Ishida, M., \& Ezuk, H., 1999, ASPC, 157, 333
\bibitem[Ishida et al.(2009)]{ish09}Ishida, M., Okada, S., Hayashi, T., et al., 2009, PASJ, 61, 77 
\bibitem[Kepler et al.(2007)]{kep07}Kepler, S. O., Kleinman, S.J., Nitta, A., et al., 2007, \mnras, 375, 1315
\bibitem[Kepler et al.(2013)]{kep13}Kepler, S. O., Pelisoli, I., Jordan, S., et al., 2013, \mnras, 429, 2934
\bibitem[Li \& Wang (2007)]{lizy07}Li, Z., \& Wang, D., 2007, \apjl,668, 39
\bibitem[Longair(2011)]{lon11}Longair, M. S., 2011, High Energy Astrophysics, Cambridge University Press, 3rd edition
\bibitem[Li \& Wang (2013)]{li13} Li, J.-T., \& Wang, Q. D. 2013, \mnras, 428, 2085
\bibitem[Mallick(1998)]{mal98}Mallik, S. V., 1998, \aap, 338, 623
\bibitem[Molaro et al. (2014)]{molaro14}Molaro, M., Khatri, R., \& Sunyaev, R. A. 2014, A\&A, 564, 107
\bibitem[Mukai et al. (2009)]{muk09}Mukai, K., Zietsman, E., \& Still, M., 2009, \apj, 707, 652
\bibitem[Mukai et al.(2012)]{muk12}Mukai, K., et al., 2012, Baltic Astronomy, 21, 54
\bibitem[Muno et al. (2007)]{muno07} Muno, M. P., Baganoff, F. K., Brandt, W. N., Park, S., \& Morris, M. R., 2007, \apj, 656, 69
\bibitem[Neustroev \& Zharikov (2008)]{neu08}Neustroev, V., \& Zharikov, S., 2008, \mnras, 386, 1366
\bibitem[Neustroev \& Tsygankov(2014)]{neu14}Neustroev, V., \& Tsygankov, S., 2014, The X-ray Universe, 294
\bibitem[Ozdonmez et al.(2015)]{ozd15}Ozdonmez, A., Ak, T., \& Bilir, S., 2015, New Astronomy, 34, 234O
\bibitem[Perez et al. (2015)]{Perez15} 	Perez, K., Hailey, C.J., Bauer, F.E., et al.  2015, Nature, 520, 646
\bibitem[Perryman et al.(1997)]{per97} Perryman, M.A.C., lindegren, L., Kovalevsky, J.,  et al., 1997, \aap, 323, L49
\bibitem[Pretorius \& Knigge(2012)]{pre12}Pretorius, M. L., \& Knigge, C., 2012, \mnras, 419, 1442
\bibitem[Pretorius \& Mukai(2014)]{pre14}Pretorius, M., \& Mukai, K., 2014, \mnras, 442, 2580
\bibitem[Rana et al.(2006)]{ran06}Rana, R., Singh, K. P., Schlegel, E. M., \& Barrett, P. E., 2006, \apj, 642, 1042
\bibitem[Reis et al.(2013)]{reis13}Reis R. C., Wheatley, P.J., Gansicke, B.T., et al., 2013, \mnras, 430, 1994	 
\bibitem[Revnivtsev et al.(2006)]{rev06} Revnivtsev, M., Sazonov, S., Gilfanov, M., Churazov, E., \& Sunyaev, R., 2006, \aap, 452, 169 
\bibitem[Revnivtsev et al.(2009)]{rev09}Revnivtsev, M., Sazonov, S., Churazov, E., et al.,  2009, \nat, 458, 1142
\bibitem[Revnivtsev et al.(2014)]{rev14}Revnivtsev, M. G., Filippova, E. V., \& Suleimanov, V. F., 2014, Astronomy Letters, V40., No. 4, 177
\bibitem[Ritter \& Kolb(2003)]{rit03}Ritter, H., \& Kolb, U., 2003, \aap, 404, 301
\bibitem[Saitou et al.(2012)]{sai12}Saitou, K., Tsujimoto, M.m Ebisawa, K., et al., 2012, PASJ, 64, 88
\bibitem[Sazonov et al.(2006)]{saz06}Sazonov, S., revnivtsev, M., Gilfanov, M., et al., 2006, \aap, 450, 117
\bibitem[Schlegel et al.(2014)]{sch14}Schlegel, E.M., Shipley, H.V., Rana, V.R., et al., 2014, \apj, 797, 38
\bibitem[Terada et al.(2010)]{ter10}Terada, Y., 2010, \apj, 721, 1908 
\bibitem[Thorstensen(2003)]{tho03}Thorstensen, J., 2003, \apj, 126, 3017
\bibitem[Thorstensen et al.(2008)]{tho08}Thorstensen, J. R., Lépine, S., \& Shara, M., 2008, \aj, 136, 2107
\bibitem[Uchiyama et al.(2011)]{uch11} Uchiyama, H., Nobukawa, M., Tsuru, T. G., Koyama, K., \& Matsumoto, H. 2011, PASJ, 63, 903
\bibitem[Uchiyama et al.(2013)]{uch13}Uchiyama, H., Nobukawa, M., Tsuru, T.,  et al., 2013, PASJ, 65, 19
\bibitem[Warwick et al.(2014)]{war14}Warwick, R. S., Byckling, K., \& Pérez-Ramírez, D., 2014, \mnras, 438, 2967	
\bibitem[Watson(1995)]{wat95}Watson, M. G., 1995, \mnras, 273, 681
\bibitem[Wijnen et al.(2015)]{wij15}Wijnen, T. P. G., Zorotovic, M., \& Schreiber, M. R., 2015, \aap, 577, 143
\bibitem[Worrall et al. (1982)]{Wor82} Worrall, D. M., Marshall, F. E., Boldt, E. A., \& Swank, J. H. 1982, ApJ, 255, 111
\bibitem[Yamauchi et al.(2009)]{yam09}Yamauchi, S., Ebisawa, K., Tanaka, Y., et al., 2009, PASJ, 61, S225
\bibitem[Yuasa et al.(2010)]{yua10}Yuasa, T., Nakazawa, K., Makishima, K., et al., 2010, \aap, 520, 25
\bibitem[Yuasa et al.(2012)]{yua12}Yuasa, T., Makishima, K., \& Nakazawa, K., 2012, \apj, 753, 129
\bibitem[Zorotovic et al.(2011)]{zor11}Zorotovic M., Schreiber M. R., \& Gansicke B. T., 2011, \aap, 536, 42




\end{thebibliography}
\end{document}